\documentclass[12pt]{iopart}
\usepackage{graphicx}
\newcommand\dslash{\slash\hspace*{-0.7em}D}
\newcommand \hmu {\hat{\mu}}
\newcommand\avr[1]{\left\langle{#1}\right\rangle}
\begin{document}

\title[Lattice QCD and heavy ion collisions: a review of recent progress]{Lattice QCD and heavy ion collisions: a review of recent progress}

\author{Claudia Ratti}

\address{Department of Physics, University of Houston, Houston, TX 77204, USA}
\ead{cratti@uh.edu}
\vspace{10pt}
%\begin{indented}
%\item[]February 2014
%\end{indented}

\begin{abstract}
In the last few years, numerical simulations of QCD on the lattice have reached a new level of accuracy. A wide range of thermodynamic quantities is now available in the continuum limit and for physical quark masses. This allows a comparison with measurements from heavy ion collisions for the first time. Furthermore, calculations of dynamical quantities are also becoming available. The combined effort from first principles and experiment allows us to gain an unprecedented understanding of the properties of quark-gluon plasma. I will review the state-of-the-art results from lattice simulations and connect them to the experimental information from RHIC and the LHC.
\end{abstract}

%\maketitle

\section{Introduction}
Strongly interacting matter undergoes a phase transition from a hadronic, confined phase, to a deconfined plasma of quarks and gluons (QGP) under extremely high temperatures or densities. Just a few microseconds after the Big Bang, the opposite transition took place; the building blocks of matter, the hadrons, were formed at this point. The same conditions of the early Universe can be re-created in the laboratory, in ultra-relativistic heavy-ion collision (HIC) experiments currently taking place at the Large Hadron Collider (LHC) at CERN, and the Relativistic Heavy Ion Collider (RHIC) at Brookhaven National Laboratory (BNL), soon to be followed by FAIR (GSI) and NICA (Dubna). While the LHC is running at the highest possible energies, thus exploring the high-temperature, zero net-density phase, the other experiments are performing a Beam Energy Scan (BES) in order to study the finite density behavior of strongly interacting matter. In particular, RHIC will run a second BES program in 2019/2020, which will focus on low collision energies with increased luminosity and enable us to reach
higher statistical precision and thus turn trends and features into definitive conclusions. Since its discovery \cite{Heinz:2000bk,Adams:2005dq,Back:2004je,Arsene:2004fa,Adcox:2004mh,Shuryak:2004cy,Gyulassy:2004zy}, it quickly became clear
that the QGP exhibits many unexpected features, being the smallest, hottest, most perfect fluid ever observed.
Quantum Chromodynamics (QCD), the fundamental theory of strong interactions, is a non-abelian gauge theory which can only be solved numerically in the strongly coupled regime of relevance in the vicinity of the phase transition. These simulations are performed by placing quarks and gluons on a discretized lattice and simulating their interactions in thermal equilibrium. There is a general consensus that the system formed in HICs is close to thermal equilibrium, which endorses lattice QCD as one of the main tools to study the underlying
thermodynamics from the theoretical point of view.

These numerical simulations are hindered by two main limitations: the sign problem does not allow simulations at finite net-baryon densities, while extracting dynamical quantities is notoriously difficult because it requires the application of inversion methods or modeling in order to integrate over a discrete set of lattice points.
There are exploratory studies in various toy models and QCD-like theories where the sign problem could be solved either by
finding appropriate dual variables \cite{Gattringer:2014nxa} or by the use of Lefschetz thimbles \cite{Alexandru:2015sua,Alexandru:2015xva,Scorzato:2015qts}. For realistic
systems, however, these techniques have not yet been generalized, and evidence is lacking that such a
generalization is possible at all. Large scale studies use
simulations at the physical point and extract physics at positive chemical potentials through analytical
continuation from imaginary chemical potential, or a Taylor expansion of thermodynamic quantities around chemical potential $\mu_B=0$. These methods enable us to explore a region of the phase diagram which is at present limited to $\mu_B/T\simeq2$. For these reasons, the QCD phase diagram in the $T$ and $\mu_B$ plane is still vastly unexplored. 
It has been shown that the transition from the hadronic to the partonic system at zero baryochemical
potential is a broad cross-over \cite{Aoki:2006we}: the change between the phases as a function of the
temperature happens gradually. One of the main unanswered questions is whether the transition between the hadronic phase and the QGP becomes first order at large net-baryonic density. Understanding the emergence of critical phenomena in the theory of strong interactions has become a cardinal challenge not only for theory but also for experiments since, depending on the location of the critical end point (CEP) in the $(T,\mu_B)$ plane, its effects may be probed in HICs \cite{Stephanov:1998dy}.

Lattice QCD simulations have reached unprecedented levels of accuracy in recent years, as for the first time they can be performed with realistic values of the physical parameters (e.g. quark masses) and on fine lattices, which leads to a very precise determination of observables that can be compared to experimental measurements. This work presents an overview of the most recent lattice results, in connection to the physics of heavy-ion collisions at RHIC and the LHC. The manuscript is organized as follows: in Section \ref{2}, a brief introduction to QCD on a discretized grid is given; Section \ref{3} is meant as the main body of this report and it contains several subsections on the most interesting observables from QCD thermodynamics: the QCD equation of state and fluctuations of conserved charges at zero and finite density, the QCD transition line and the partial pressures of hadronic states divided into families according to their flavor and baryon number content. The comparison to experimental observables and its limitations are discussed. Section \ref{4} presents results on transport coefficients of experimental relevance such as shear viscosity, electric conductivity and heavy flavor diffusion coefficient. In Section \ref{5}, available results on the spectral functions of heavy quarkonia and the identification of the charm degrees of freedom in the QGP through fluctuations are discussed. Section \ref{6} contains conclusions and outlook.

%{\bf General introduction on the physics of heavy ion collisions and quark-gluon plasma. First mentioning of the issues of lattice QCD simulations (e.g. sign problem and dynamical quantities). Outline of the various sections of the manuscript.}
\section{QCD thermodynamics on a discretized grid \label{2}}
Lattice field theory is based on a path-integral representation of the theory: the QCD partition function can be written in terms of a Euclidean path integral over quark (Dirac spinor $\bar{\psi}_f,\psi_f$) and gluon (SU(3) vector $A_\mu$) fields. Such partition function will depend on the temperature $T$, volume $V$, quark masses $m_f$ and chemical potentials $\mu_f$. The index $f$ labels the different quark flavors. In order to apply powerful techniques developed
in statistical mechanics, the theory is considered in the Euclidian space-time by substituting the real time $t$
with an imaginary time, $t\rightarrow-ix_4$, with real $x_4$. The lattice discretization of the Euclidian space-time defines the lattice field theory
we shall study. We will consider mainly 4-dimensional hyper-cubic lattices $N_s^3\times N_t$ with three spatial dimensions $N_s$ and a temporal dimension $N_t$, unless stated otherwise. The temperature is defined as the inverse of the temporal extension of the lattice: $T=1/(N_t*a)$, where $a$ is the lattice spacing. The lattice points are called sites and the bonds connecting the nearest neighbor sites are called links. Matter fields are defined on the sites, while gauge fields live on the links $U_\mu(x)=\exp(iaA_\mu(x))$. The simplest way of discretizing the theory is by replacing the derivatives in the continuum euclidean action by finite differences. This trivial action can be improved in order to reduce the discretization effects, e.g. by including further differences in the derivative than just the nearest neighbors.

For fermions on the lattice, a complication exists in the formulation: in addition to the physical mode, there are so called ``doublers" which arise as a consequence of the fact that the derivative in the kinetic term of the Dirac action is first order. These unwanted modes survive as relevant degrees of freedom in the continuum limit. Staggered fermions are constructed in order to partially reduce the number of doublers: for each physical quark flavors, there are four in the Staggered formulation. The flavor-chiral symmetry is broken due to a flavor-mixing interaction at finite lattice spacing. However, at least a part of the chiral symmetry is preserved. This makes a numerical analysis of chiral properties much easier than in the case of Wilson fermions, and staggered fermions require less computer resources than Wilson fermions as well. The latter are constructed by introducing a second-derivative term in the action; in this way, the doublers are decoupled in the limit $a\rightarrow0$. However, because the Wilson term is essentially the mass term for doublers, chiral symmetry is violated even in the limit of vanishing bare quark mass.

The finite temperature field theory is defined by the Matsubara formalism for finite temperature statistical systems. We consider static systems in thermal equilibrium at temperature
$T$. The partition function of QCD in the Euclidean space-time is given by:
\begin{equation}
Z(T,V)=\int\prod_{\mu} \mathcal{D}A_\mu\prod_{f} \mathcal{D}\psi_{f}\mathcal{D}\bar{\psi}_{f}\exp(-S_E(T,V))
\end{equation}
where 
\begin{eqnarray}
S_E(T,V)=-\int_0^{1/T}dx_4\int_V d^3x \mathcal{L}^E,
\nonumber\\
\mathcal{L}^E=-\sum_f\bar{\psi}_f(x)(\,\slash\hspace{-.28cm}{\mathcal{D}}+m_f)\psi_f(x)-\frac14F_{\mu\nu}^a(x)F^{\mu\nu}_a(x)
\end{eqnarray}
and
\begin{equation}
\slash\hspace{-.28cm}{\mathcal{D}}=(\partial_\mu+ig\frac{\lambda^a}{2}A_{\mu}^{a})\gamma_\mu^E~~~~~~F_{\mu\nu}^a=\partial_\mu A_\nu^a-\partial_\nu A_\mu^a-gf^{abc}A_\mu^bA_\nu^c.
\end{equation}

We will be presenting results for thermodynamic quantities such as pressure $p$, energy density $\epsilon$, entropy density $s$ and trace anomaly $I$. These quantities are defined as derivatives of the partition function with respect to the temperature and physical volume of the system:
\begin{eqnarray}
p=T\frac{\partial \ln Z}{\partial V}~~~~~~~~~~~~~~~~~~\epsilon=-\frac{1}{V}\frac{\partial \ln Z}{\partial T^{-1}}
\nonumber\\
I=\epsilon-3p~~~~~~~~~~~~~~~~~~~~~s=\frac{\epsilon+p}{T}.
\end{eqnarray}
The thermal expectation value of physical observables $\mathcal{O}$ can be expressed as
\begin{equation}
\langle\mathcal{O}\rangle=\frac{1}{Z(T,V)}\int\prod_{\mu} \mathcal{D}A_\mu\prod_{f} \mathcal{D}\psi_{f}\mathcal{D}\bar{\psi}_{f}\mathcal{O}\exp(-S_E(T,V)).
\end{equation}

In principle, the above listed quantities also depend on the quark chemical potentials. However, as already mentioned, the fermion sign problem prevents the numerical calculation of thermodynamic quantities at $\mu_i\neq0$. A possible way to circumvent this problem is through analytically continued results of simulations at imaginary chemical potentials, which we briefly explain here as it will be useful later in the text.
The chemical potential is introduced through
weighted temporal links in the staggered formalism:
\begin{eqnarray}
U_0(\mu)=e^\mu U_0,~~~~~~
U_0^\dagger(\mu)=e^{-\mu} U_0^\dagger.
\end{eqnarray}
Thus, an imaginary $\mu$ translates into a phase factor for the
antiperiodic boundary condition in the Dirac operator. Due to the $Z(3)$
symmetry of the gauge sector, there is a non-trivial periodicity in
the imaginary chemical potential $\mu_q \to \mu_q+i(2\pi/3)T$, which translates
to the baryochemical potential as $\mu_B \to \mu_B+i 2\pi T$, the
Roberge-Weiss symmetry. This is independent of the charge conjugation symmetry 
$\mu_B \leftrightarrow -\mu_B$. 

The introduction of the imaginary chemical potential does not break
the $\gamma_5$-Hermiticity of the Dirac operator, it can simply be introduced
as a phase shift in the time-like boundaries. The chemical potential can
be implemented on a flavor-by-flavor basis. One can then have different choices for the chemical potentials $\mu_i$. Usually, $\mu_u=\mu_d$, while for the strange chemical potential different possibilities arise. A non-trivial choice is the tuning to $\langle n_S\rangle=0$, useful for phenomenological purposes.
This requires the solution of a differential equation, where the $\mu_S^I(\mu_B^I)$ function is to be determined:

\begin{equation}
\frac{d}{d\mu_B^I} \frac{\partial \log Z}{\partial \mu_S} =0\,.
\end{equation}
with the trivial initial condition $\partial\ln Z/\partial\mu_S=0$ at $\mu_B^I=0$.

%Short theoretical introduction on how to solve the theory of strong interactions on a lattice.
\section{Bulk properties of QCD matter \label{3}}
The most reliable and precise results available today from lattice QCD simulations concern the calculation of thermodynamic observables in equilibrium. Due to a steady and continuous improvement in computer resources, numerical algorithms and our physical understanding
which manifests itself in physical techniques (e.g. the Wilson-flow scale setting introduced in
Ref. \cite{Borsanyi:2012zs}), final results for several quantities exist in the continuum limit and for physical values of the parameters such as the quark masses. Therefore, for the first time they can be reliably utilized in a variety of applications:
\begin{enumerate}
\item{as inputs in the phenomenological modeling of heavy ion collisions such as hydrodynamic simulations;}
\item{to test models which try to identify the effective degrees of freedom at a given temperature and chemical potential;}
\item{to tune models which are then used to extend the lattice approach to large density or to the calculation of dynamical quantities;} 
\item{to test the range of applicability of perturbation theory;}
\item{in comparison to experimental measurements, in order to extract the properties of the system created in heavy ion collisions from first principles.}
\end{enumerate}
To zeroth order in the coupling, all thermodynamic quantities can be evaluated analytically. 
Since the effective QCD coupling goes to zero logarithmically at short distances, it is reasonable to attempt a perturbative expansion of the thermodynamic potential at high energy density. A perturbation series can be constructed and truncated at a given order in the expansion parameter (the QCD coupling constant). The convergence of this series turns out to be slow; for this reason, resummation techniques (such as Hard Thermal Loop) or dimensional reduction have been proposed and successfully applied to several observables. Deviations of the lattice QCD results at large temperatures from the ideal gas limit are observed, much more prominently in global thermodynamic observables than in quantities which do not involve the gluonic contribution at tree level, such as fluctuations of conserved charges. This shows that the gluonic interaction is still conspicuous at large temperatures. It is important to test at which temperature the perturbative approach will break down, which in general depends on the observable under study.

In the low temperature phase, the results of lattice simulations for QCD thermodynamics are well described by the Hadron Resonance Gas (HRG) model, whose roots are in the theorem by Dashen,
Ma and Bernstein \cite{Dashen:1969ep} which allows one to calculate the microcanonical partition function of
an  interacting  system, in the thermodynamic limit $V\rightarrow\infty$, assuming that it is a gas of non-interacting free hadrons and resonances \cite{Venugopalan:1992hy}.

In between these two opposite regimes of temperatures, lattice QCD simulations are needed, to calculate thermodynamic quantities from first principles. In this section, an overview of the most recent results will be provided.
%{\bf This section, meant as the main body of the report, contains several subsections on the most interesting observables from QCD thermodynamics. These are the quantities for which the most advanced lattice QCD results are available, and for some of them it is possible to have a quantitative comparison with experimental measurements.}
\subsection{Equation of state at $\mu_B=0$ \label{3.1}}
\begin{figure}[hb]
\begin{center}
\includegraphics[width=3in]{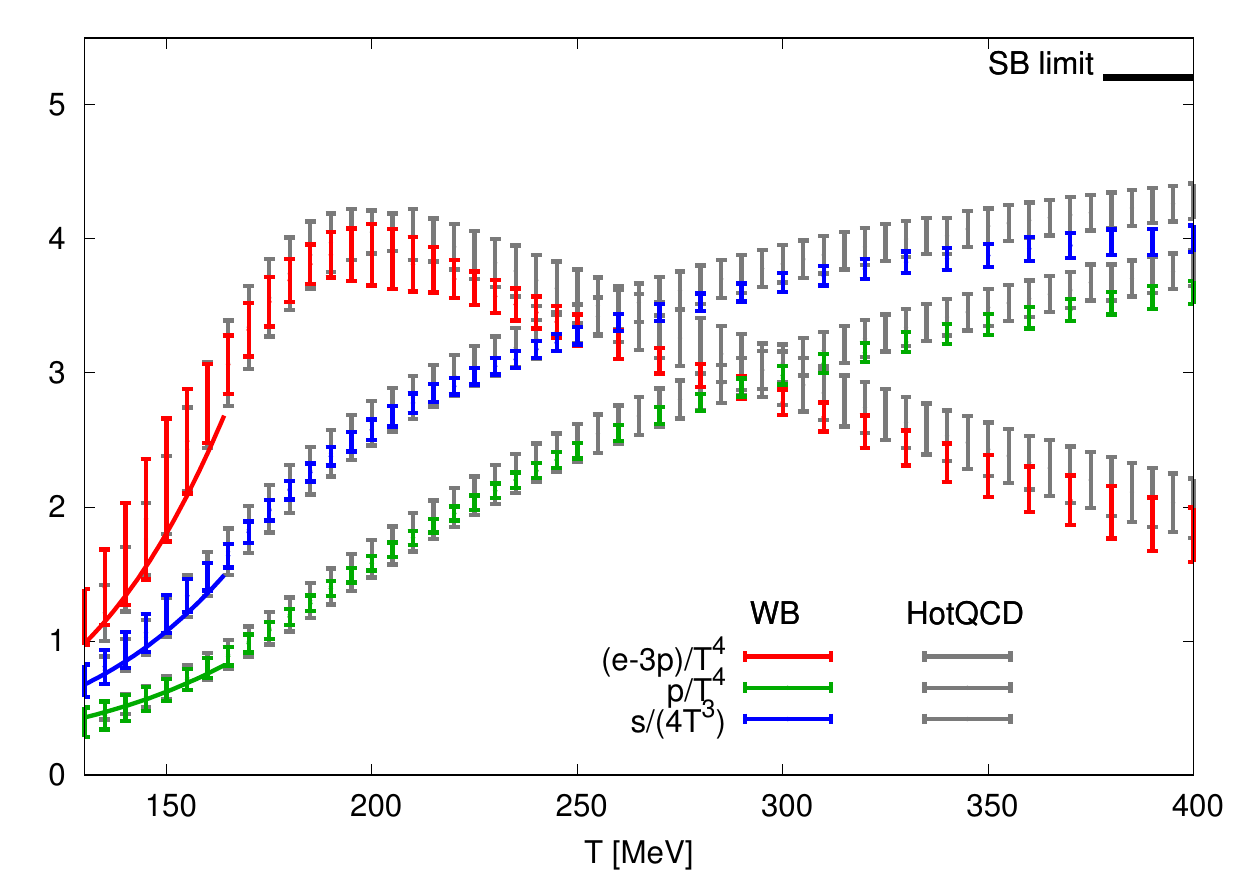}
\includegraphics[width=3in]{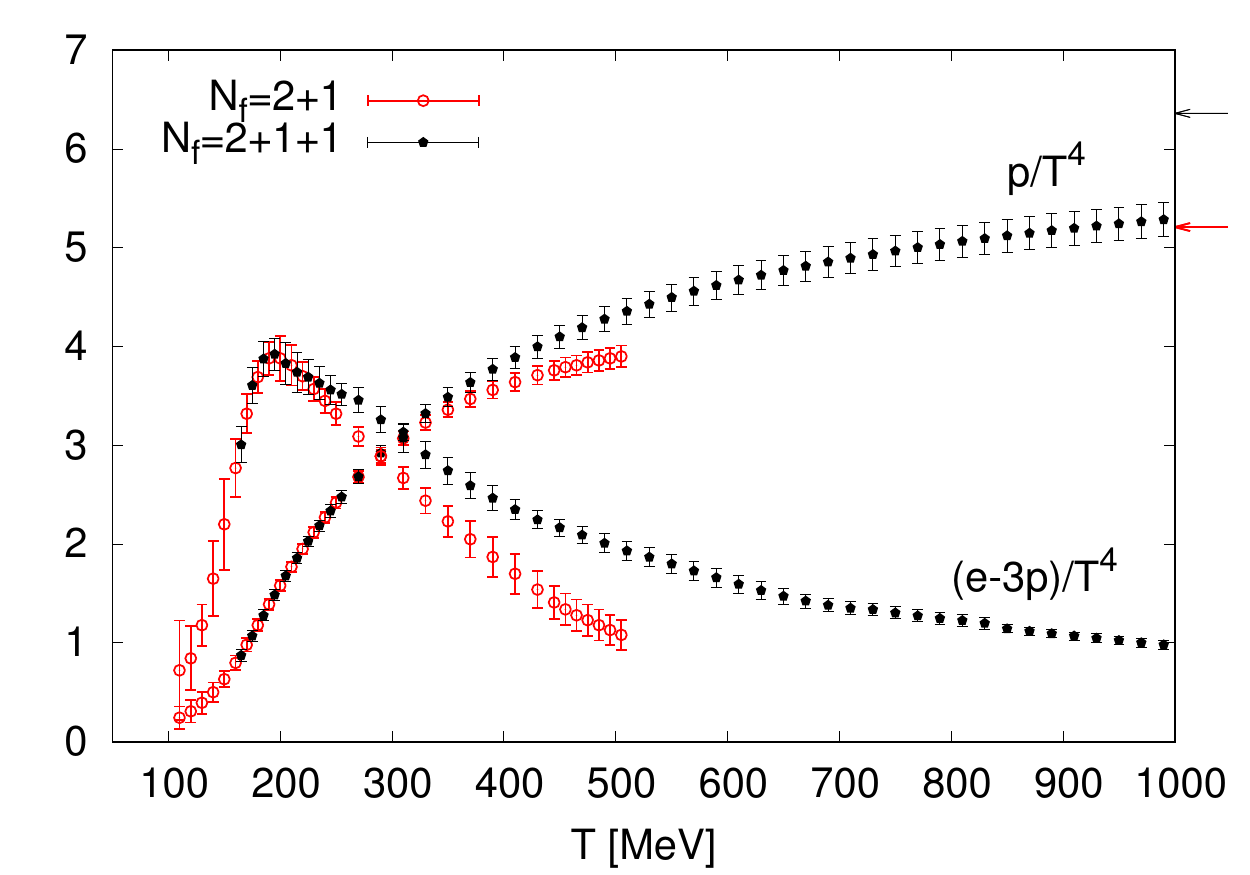}
\end{center}
\vspace{-.7cm}
\caption{\label{fig1}
Left: Continuum extrapolated results for trace anomaly, entropy density and pressure. The gray points are from the HotQCD collaboration \cite{Bazavov:2014pvz}, while the colored ones are from the WB collaboration \cite{Borsanyi:2013bia}. The figure also shows the Stefan-Boltzmann limit for the pressure and the scaled entropy; the curves at low temperature correspond to the HRG model predictions. Right: the trace anomaly and pressure in the 2+1 and 2+1+1 flavor theories (from Ref. \cite{Borsanyi:2016ksw}). 
}
\end{figure}

The Equation of State (EoS) of a system is the most relevant quantity to describe its behavior in the given physical conditions. In the case of zero net-baryonic density, the EoS of QCD is known with high precision from first principles since a few years: in 2014 the HotQCD collaboration published
continuum results for pressure, energy density, entropy density and interaction measure as functions
of the temperature \cite{Bazavov:2014pvz} which agree with the ones previously obtained by the Wuppertal Budapest (WB) collaboration \cite{Borsanyi:2010cj,Borsanyi:2013bia}. Both results have been obtained for a system of 2+1 quark flavors with physical values of the quark masses and in the continuum limit. In both cases, staggered fermions have been used, but with two different discretizations: the WB collaboration used the tree-level Symanzik improved gauge, and stout-improved staggered fermion action with two levels of smearing (2stout) \cite{Aoki:2005vt}, while the HotQCD collaboration used the highly improved staggered quark (HISQ) action introduced in \cite{Follana:2006rc}.
The fact that the two analyses, which differ at finite lattice spacing, yield the same result in the continuum limit is a fundamental test of the lattice approach to QCD thermodynamics. The continuum extrapolated results for interaction measure, entropy density and pressure are shown in the left panel of Fig. \ref{fig1}. The gray points are from the HotQCD collaboration, while the colored ones are from the WB collaboration. The figure also shows the Stefan-Boltzmann limit for the pressure and the scaled entropy; the curves at low temperature correspond to the HRG model predictions. Notice that, as already anticipated, the thermodynamic quantities are still relatively far from the corresponding ideal gas value, even at temperatures of the order of 400 MeV. Some observables show a good agreement with resummation \cite{Andersen:2011sf} or dimensional reduction \cite{Laine:2006cp} techniques already at $T\sim400 $ MeV, while for others this is not the case. For example, in the case of the trace anomaly, the results from Hard Thermal Loop perturbation theory to three-loop order \cite{Andersen:2011sf} show a large uncertainty, corresponding to varying the renormalization scale. The results from Electrostatic QCD are in good agreement with the HTL ones, and the lattice results approach the perturbative ones at $T\sim 2-3T_c$.
%\vspace{-.2cm}
%\vspace{-.2cm}

More rigorous formulations (e.g. simulations with Wilson, overlap or domain wall fermions) will serve as important cross-checks in the coming years. It is worth mentioning that, recently, first results for the
equation of state obtained from other approaches to lattice QCD are becoming available: these include the
gradient flow method \cite{Asakawa:2013laa,Kanaya:2016rkt}, which extracts the thermodynamic quantities from the energy-momentum tensor,
and twisted mass fermions \cite{Burger:2014xga}; the former are limited so far to the quenched approximation \cite{Asakawa:2013laa} or heavier than physical quark masses \cite{Kanaya:2016rkt}, the latter to two flavors with heavier-than-physical quark masses.

More recently, the effect of the charm quark on thermodynamic quantities has been investigated \cite{Borsanyi:2016ksw}. From this analysis, it turns out that the charm quark is a relevant degree of freedom already at $T\sim250$ MeV, which should be taken into account in hydrodynamic simulations of heavy ion collisions at the LHC energies \cite{Alba:2017hhe}. This is shown in the right panel of Fig. \ref{fig1}.
\begin{figure}[hb]
\begin{center}
\includegraphics[width=5.7in]{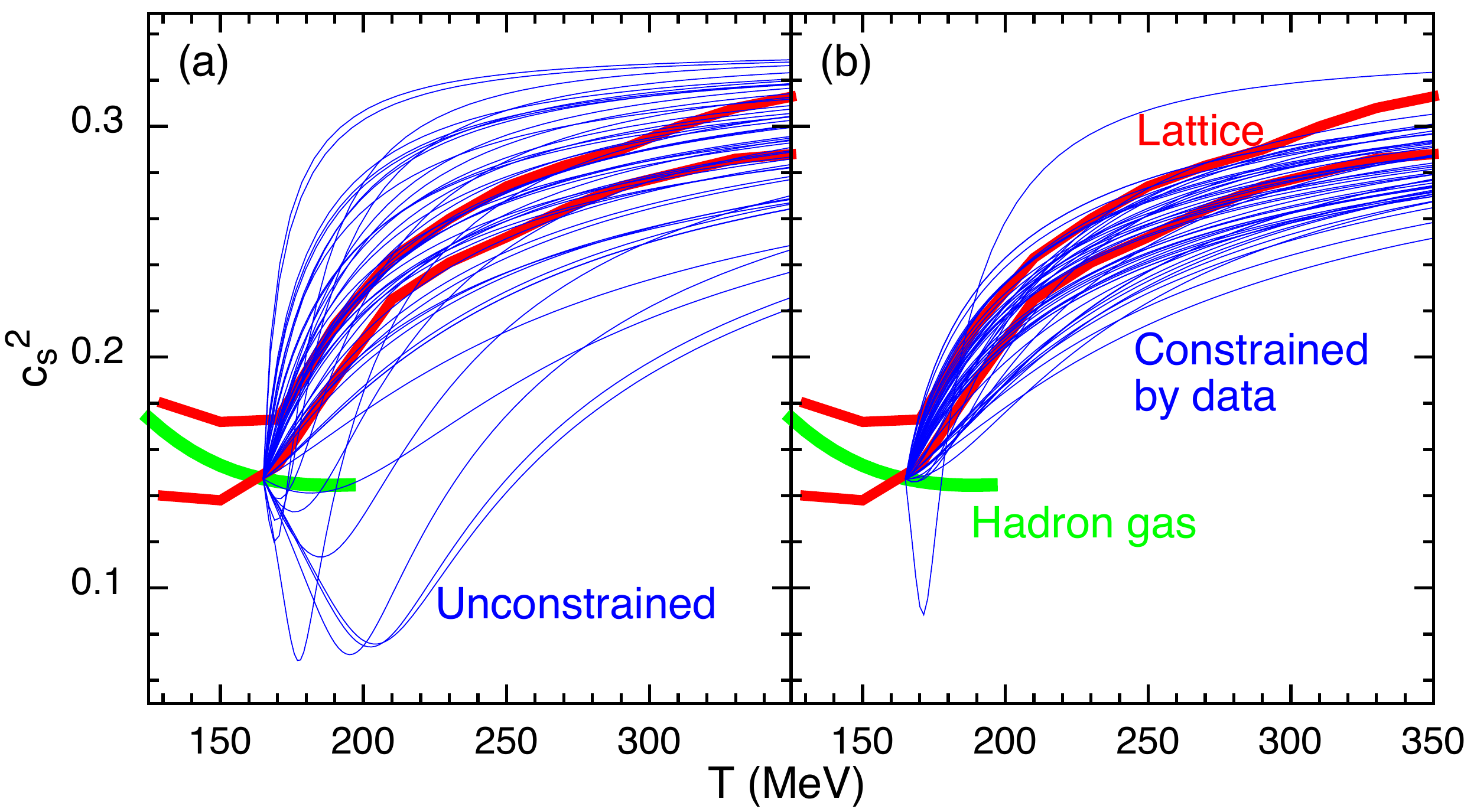}
\end{center}
%\vspace{-.5cm}
\caption{\label{fig2}
From Ref \cite{Pratt:2015zsa}: Constraints on the QCD equation of state from the Bayesian analysis. (a) Fifty equations of state were generated by randomly choosing
the parameters from the prior distribution and weighted by the posterior likelihood (b).  The two
red  lines  in  each  figure  represent  the  range  of  lattice equations of state shown in \cite{Bazavov:2014pvz}, and the green line
shows the equation of state of a non-interacting hadron gas.
This suggests that the matter created in heavy-ion collisions
at RHIC and at the LHC has a pressure that is similar to that expected from equilibrated matter.
}
\end{figure}

Recently, an important validation of the lattice QCD Equation of State has been obtained from a Bayesian analysis \cite{Pratt:2015zsa}. This framework, based on a comparison of data from RHIC and the LHC to theoretical models, has applied state-of-the-art statistical techniques to the combined analysis of a large number of observables while varying the model parameters. The posterior distribution over possible equations of states turned out to be consistent with results from lattice QCD simulations, as shown in Fig. \ref{fig2}. This analysis has also been successfully applied to infer the behavior of other quantities, such as the shear viscosity of the QGP at zero \cite{Bernhard:2016tnd} and finite density \cite{Auvinen:2017fjw}.
%{\bf Summary of the state-of-the-art results on the QCD Equation of State at zero density. Continuum extrapolated results from the Wuppertal-Budapest and HotQCD collaborations with staggered fermions. Mention of other approaches, such as the gradient flow method and twisted mass fermions. Cite the bayesian validation!!!}

It is worth pointing out that results exist for the equation of state of QCD in background magnetic fields \cite{Bonati:2013vba,Bali:2014kia}: in Ref. \cite{Bali:2014kia} the equation of state for a system of 2+1 flavors at physical quark masses has been obtained, together with the magnetic susceptibility and permeability, which show that strongly interacting matter is paramagnetic around and above the transition temperature.
\subsection{Equation of state at $\mu_B\neq0$}
The equation of state of strongly interacting matter at finite density is a very relevant quantity, among other things, for the low energy runs of heavy ion collisions and for neutron star physics.  It is worth mentioning that recently, results from perturbative QCD at very large density have been obtained and used to constrain neutron star matter \cite{Fraga:2013qra}. Extracting the equation of state (and other properties) of QCD at finite chemical potential from regular Monte Carlo simulations is not possible at the moment. Indeed, ab initio calculations in the baryon dense regime of QCD are hindered by the fermion sign problem, a fundamental technical obstacle of exponential complexity \cite{Troyer:2004ge} inherent to any path integral representation of Fermi systems at finite density. 

Over the last few years, alternative methods have been proposed to extract the properties of QCD matter at small chemical potential. These include Taylor expansion around
$\mu_B=0$ \cite{Allton:2002zi,Allton:2005gk,Gavai:2008zr,Basak:2009uv,Kaczmarek:2011zz}, analytic continuation from imaginary
$\mu_B$ \cite{deForcrand:2002hgr,DElia:2002tig,DElia:2002tig,Wu:2006su,DElia:2007bkz,Conradi:2007be,deForcrand:2008vr,DElia:2009pdy,Moscicki:2009id}, reweighting of the generated configurations \cite{Fodor:2001au,Fodor:2001pe,Csikor:2004ik,Fodor:2004nz}, use of the canonical ensemble \cite{Alexandru:2005ix,Kratochvila:2005mk,Ejiri:2008xt} and density of state methods \cite{Fodor:2007vv,Alexandru:2014hga}. Here we will focus on the first two.

The pressure of QCD can be expanded in a Taylor series around $\mu_B=0$ in the following way
\begin{equation}
\hspace{-1.5cm}
\frac{p(T,\mu_B)}{T^4}=\frac{p(T,0)}{T^4} +\sum_{n=1}^\infty\left.\frac{1}{(2n)!}\frac{\mathrm{d^{2n}}(p/T^4)}{d(\frac{\mu_B}{T})^{2n}}\right |_{\mu_B=0}\left(\frac{\mu_B}{T}\right)^{2n}=\sum_{n=0}^{\infty}c_{2n}(T)\left(\frac{\mu_B}{T}\right)^{2n}.
\end{equation}
The coefficients $c_i(T)$ of the Taylor series are simulated on the lattice, either directly at $\mu_B=0$ or by using the analytical continuation technique from imaginary $\mu_B$. This means that the method traditionally used at $\mu_B= 0$
can be generalized to any imaginary $\mu_B$, and the $\mu_B$-dependence of the direct derivative is then analyzed, in order to extract higher order coefficients.
More in detail, in the direct method a derivative of the partition function can be written in terms of the action with all fermionic degrees of freedom already integrated out, $S_{eff}$, as follows:
\begin{equation}
\partial_i \log Z = \frac{1}{Z} \int{\cal D}U \partial_i e^{-S_{\rm eff}} = \avr{A_i}.
\end{equation}

Here $i$ indicates the variable of the derivative, the chemical potential $\mu_i$
in this case. $A_i$ is the first derivative of $S_{\rm eff}$ without the factor $e^{-S_{\rm eff}}$. Its
ensemble average is calculated with the same weight used for generating the
configurations. In particular,
\begin{eqnarray}
A_i&=&\frac14\textrm{tr}  M_i^{-1}(m_i,\mu_i) M_i'(m_i,\mu_i)\,,\label{eq:A}
\label{eq:B}
\end{eqnarray}
where $M_i(m_i,\mu_i)=m_i+\dslash(\mu_i)$ is the fermion operator with the bare mass $m_i$;  $M_i'(m_i,\mu_i)$ stands for
its first derivative with respect to $\mu_i$. Higher order derivatives can be evaluated in a similar way. The most expensive part of this method is the calculation of the trace
in Eq.~(\ref{eq:A}), which contains disconnected contributions and appears in
almost all susceptibilities.

After the early results for $c_2,~c_4$ and $c_6$ \cite{Allton:2005gk}, the first continuum extrapolated results for $c_2$ were published in Ref. \cite{Borsanyi:2012cr}; in Ref. \cite{Hegde:2014sta} $c_4$ was shown, but only at finite lattice spacing. The continuum limit for $c_6$ was published for the first time in \cite{Gunther:2016vcp}, and later in \cite{Bazavov:2017dus}. In \cite{DElia:2016jqh}, a first determination of $c_8$, at two values of the temperature and $N_t=8$ was presented.

\begin{figure}[h]
\begin{center}
\includegraphics[width=6.1in]{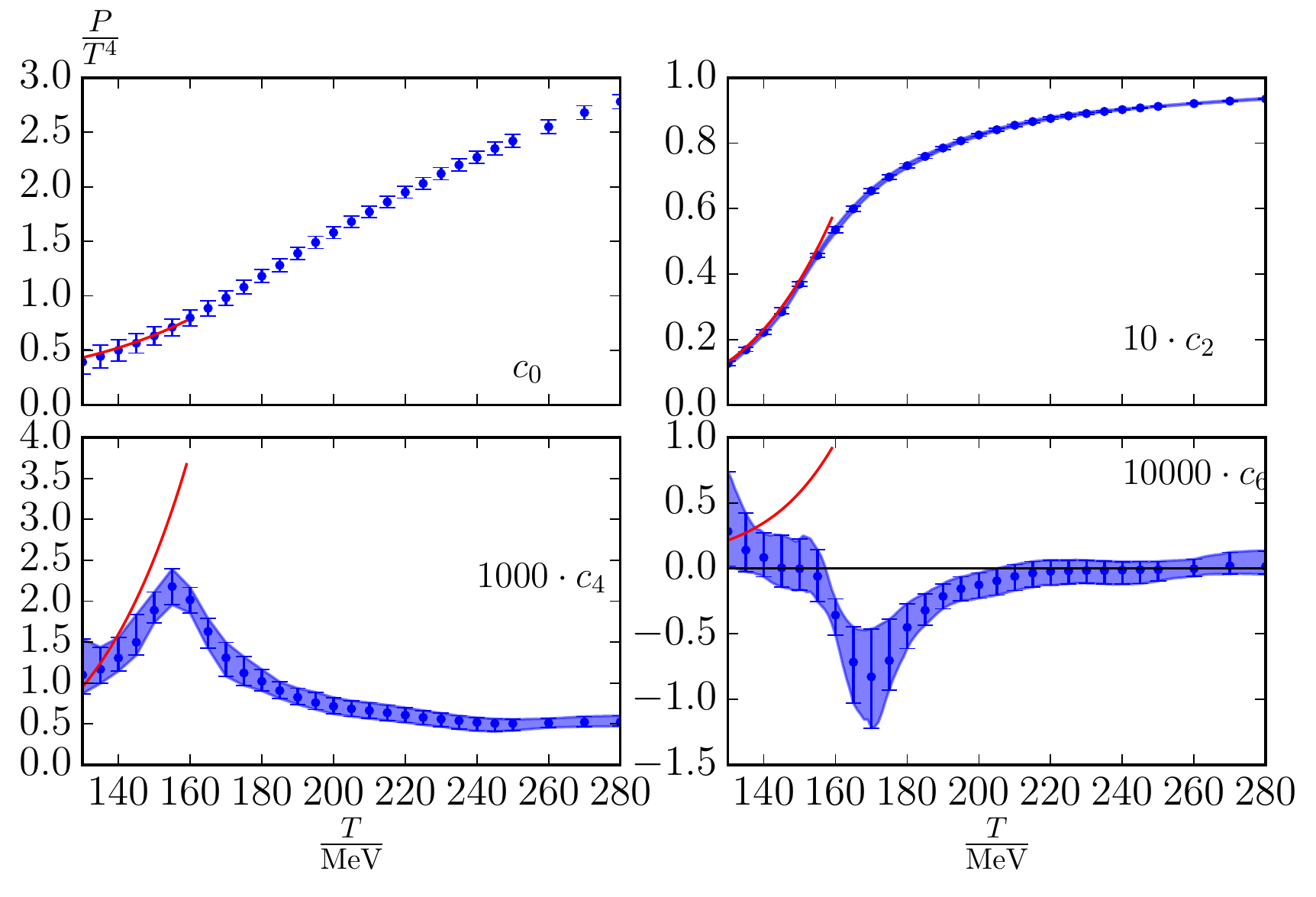}
\end{center}
%\vspace{-.5cm}
\caption{\label{fig3}
From Ref. \cite{Gunther:2016vcp}. Coefficients $c_0,...c_6$ for the Taylor expansion of the pressure
around $\mu_B=0$ in the case of strangeness neutrality. The data are continuum extrapolated and they are presented as
functions of the temperature along with the HRG prediction (red
lines).
}
\end{figure}

In Fig. \ref{fig3}, the continuum extrapolated results for $c_0,~c_2,c_4$ and $c_6$ from Ref. \cite{Gunther:2016vcp} are shown as functions of the temperature. In this case, strangeness neutrality $\langle n_s\rangle=0$ and electric charge conservation $\langle n_Q\rangle=0.4\langle n_B\rangle$ are enforced to match the experimental situation, and the coefficients are intended as total derivatives. They have been obtained by simulating the following quantity at zero and imaginary $\mu_B$:
\begin{equation}
\frac{n}{\mu_BT^2}=\frac{T}{\mu_B}\left.\frac{d(p/T^4)}{d(\mu_B/T)}
\right|_{
\langle n_S\rangle=0, ~\langle n_Q\rangle=0.4\langle n_B\rangle, T=\mathrm{const}
}.
\label{eq:density}
\end{equation}
Note that $n$ is related to the baryon number density as $\frac{n}{n_B}=1+0.4 \frac{d\mu_Q}{d\mu_B}$.
The first few terms in its Taylor expansion are: $2 c_2 + 4 c_4(\mu_B/T)^2 + 6
c_6(\mu_B/T)^4$: taking derivatives of $n/(\mu_BT^2)$ with respect to
$\mu_B$, one can obtain the desired Taylor coefficients. It turns out
that this method allows a more precise determination of these quantities,
compared to the direct simulation at $\mu_B=0$. A similar approach, based on the simultaneous fit of several observables at imaginary chemical potential, was used in Ref. \cite{DElia:2016jqh}.

\begin{figure}[ht]
\begin{center}
\includegraphics[width=3.05in]{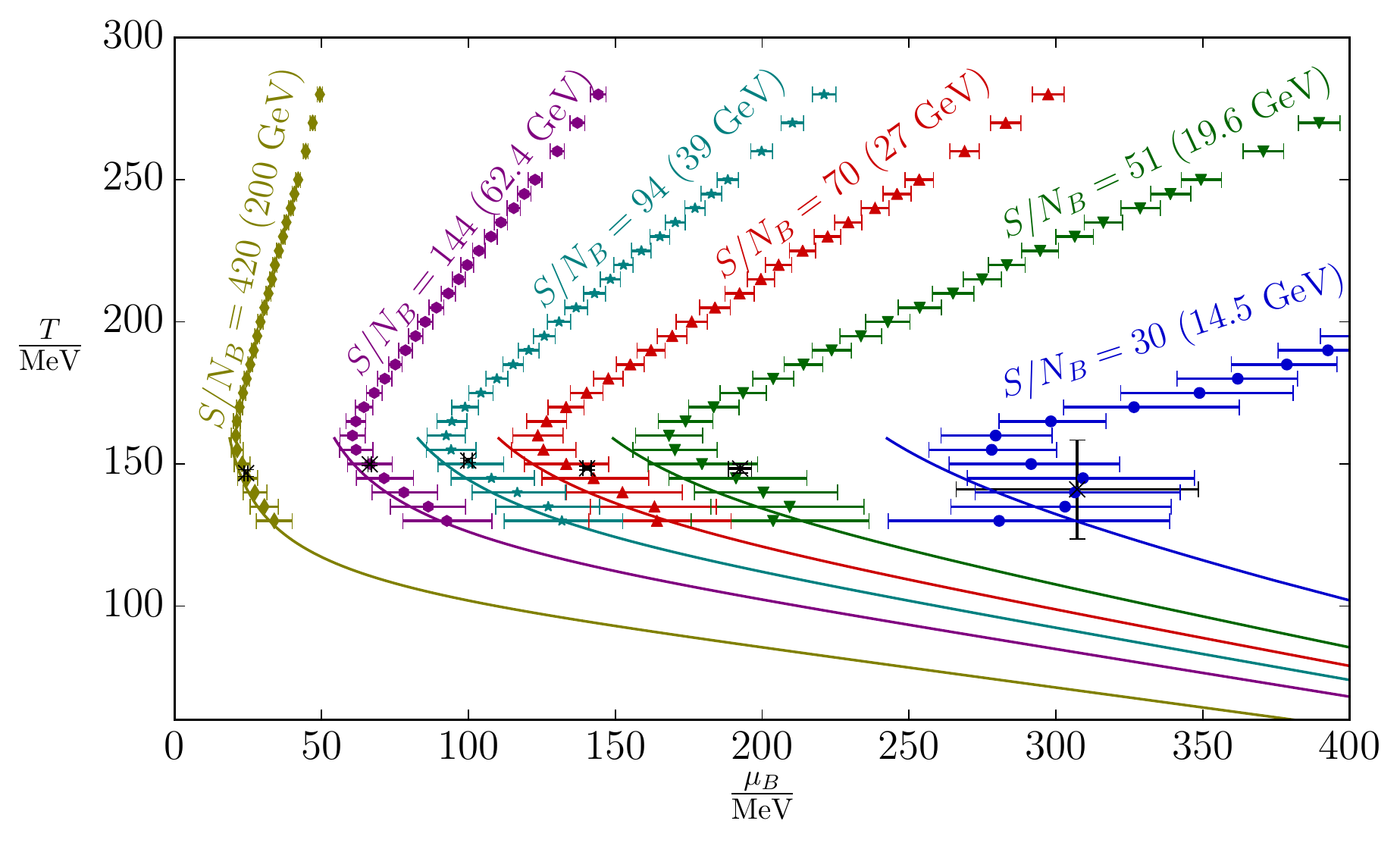}
\includegraphics[width=3.05in]{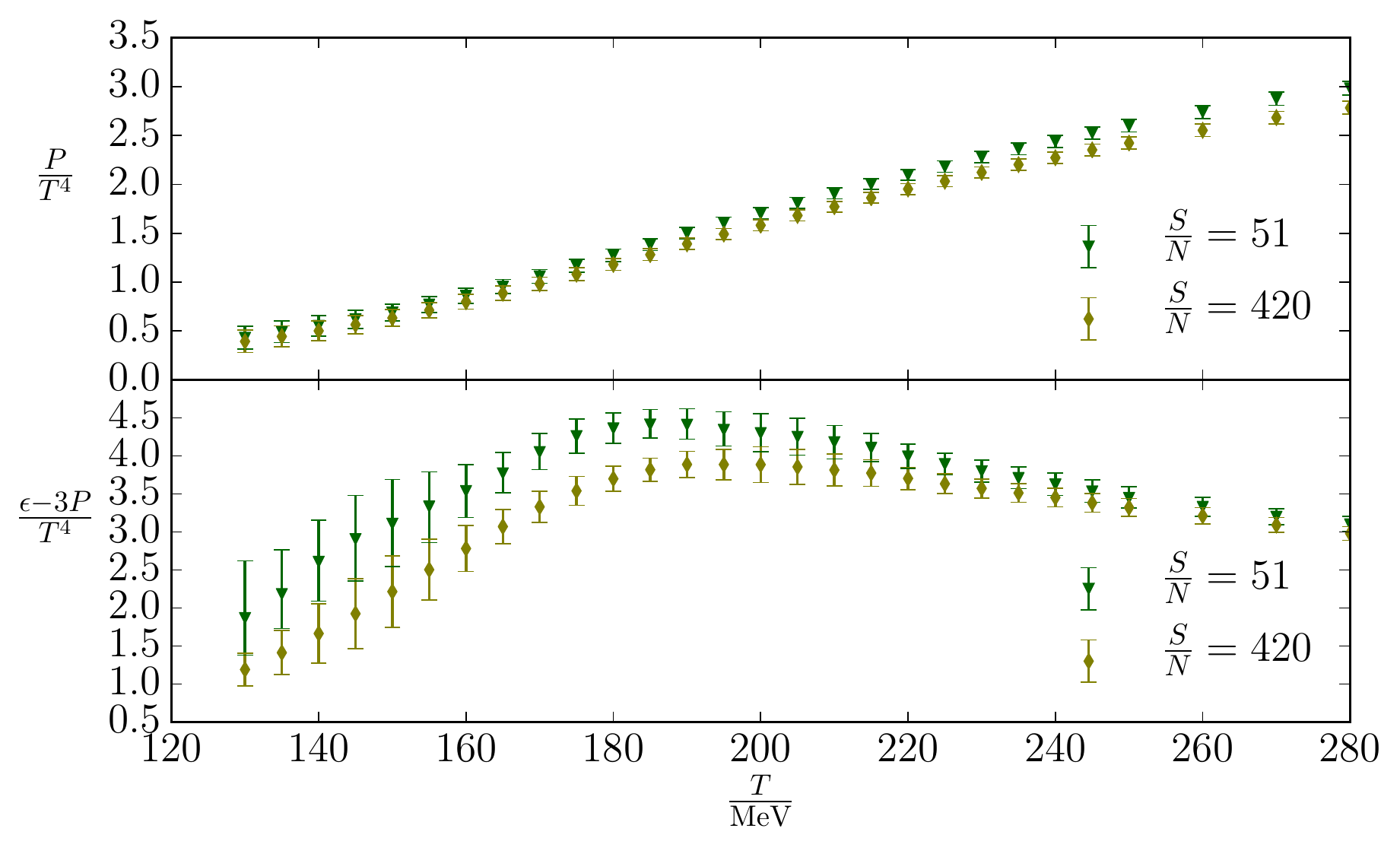}
\end{center}
%\vspace{-.5cm}
\caption{\label{fig3a}
From Ref. \cite{Gunther:2016vcp}. Left: isentropic trajectories, in the $(T,\mu_B)$ plane, that the system created in heavy ion collisions would follow in the absence of dissipation. The lines at low temperature correspond to the HRG model result. The black points correspond to the freeze-out data of Ref. \cite{Alba:2014eba}. Right: pressure and trace anomaly, calculated along two of the isentropes.}
\end{figure}

The left panel of Fig. \ref{fig3a} shows the isentropic trajectories in the $(T,\mu_B)$ plane that the system in a heavy-ion collision would follow in the absence of dissipation, namely for vanishing shear viscosity. The right panel of Fig. \ref{fig3a} shows the pressure and trace anomaly calculated along two such isentropes. Both panels are from Ref. \cite{Gunther:2016vcp}

\begin{figure}[ht]
\begin{center}
\includegraphics[width=3.05in]{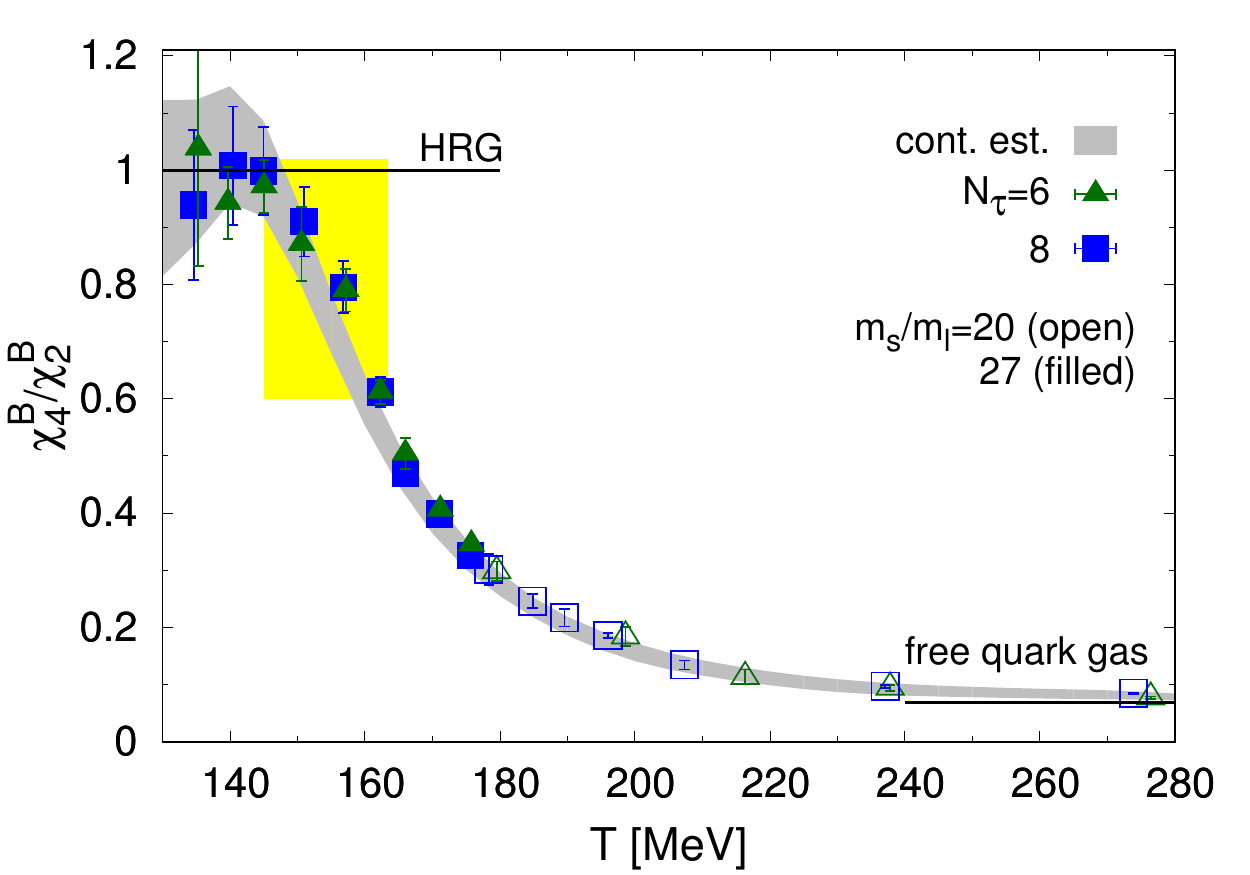}
\includegraphics[width=3.05in]{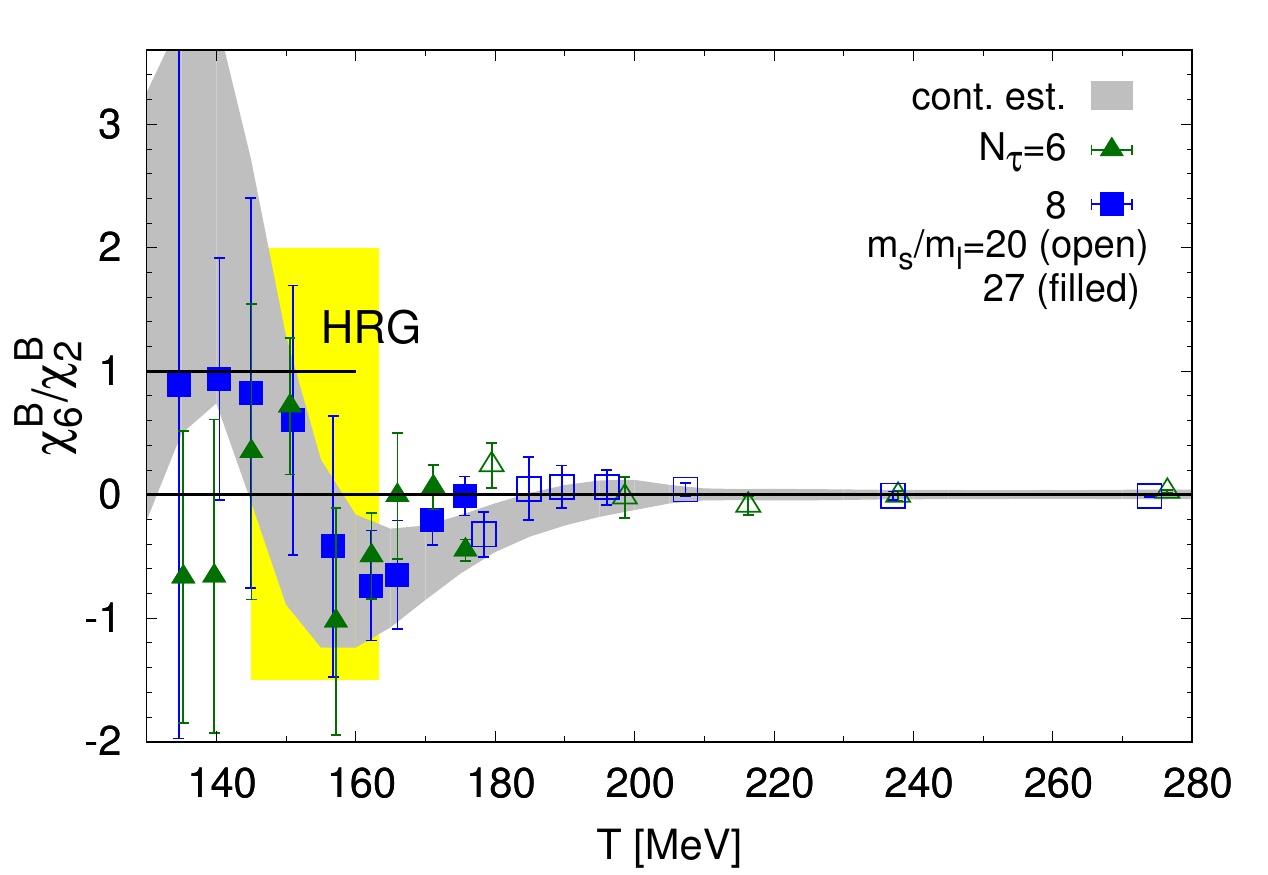}
\end{center}
%\vspace{-.5cm}
\caption{\label{fig4}
From Ref. \cite{Bazavov:2017dus}. Left: The ratio of fourth and second order Taylor expansion coefficient ($\chi^B_4/\chi^B_2$) in the case in which $\mu_S=\mu_Q=0$ as a function of the temperature ($\chi^B_n=(n!)c_n$ in this case).
Right: same as the left hand side, but for the ratio of sixth and second order cumulants of net-baryon number fluctuations
($\chi^B_6/\chi^B_2$). The boxes indicate the transition region, $T_c = (154\pm9)$ MeV. Grey bands show the continuum estimate.
}
\end{figure}

The direct method has been used in Ref. \cite{Bazavov:2017dus}, which presents a detailed extraction of the equation of state at finite $\mu_B$, up to $\mathcal{O}(\mu_B^6)$, both in the case of strangeness neutrality and for $\mu_S=\mu_Q=0$. The ratios of coefficients $c_4/c_2$ and $c_6/c_2$ corresponding to the latter case are shown in Fig. \ref{fig4}. The authors of Ref. \cite{Bazavov:2017dus} also calculated thermodynamic quantities at fixed $\mu_B/T$ up to a certain power of $\mu_B$. Recently, a holographic approach to strongly interacting matter \cite{Critelli:2017oub} has been constructed to reproduce QCD thermodynamics at $\mu_B=0$, and then extended to high density which the lattice simulations cannot reach at the moment; all the Taylor expansion coefficients available from lattice QCD are reproduced by this approach. Ref. \cite{Critelli:2017oub} contains a test of the validity of the Taylor series truncation, which shows that the full and truncated results up to $\mathcal{O}(\mu_B^6)$ from the holographic approach agree with each other up to $\mu_B/T\simeq2.5$, while for larger chemical potentials, the term $\mathcal{O}(\mu_B^8)$ is required. This shows the need of calculating higher order coefficients on the lattice, in order to have an EoS from first principles which covers the whole energy scan range at RHIC.

%\begin{figure}[ht]
%\begin{center}
%\includegraphics[width=3in]{figure5a.pdf}
%\includegraphics[width=3in]{figure5b.pdf}
%\end{center}
%\vspace{-.5cm}
%\caption{\label{fig5}
%The $\mu_B$-dependent contribution to the pressure (left) and the baryon density (right) as functions of
%$T$ for different values of $\mu_B/T$. The solid curves correspond to the full holographic result computed using black hole engineering.
%The bands denote the holographic results reconstructed through a power series expansion up to different orders in $\mu_B/T$ \cite{Critelli:%2017oub}. The points correspond to the reconstructed Taylor series up to
%$\mathcal{O}(\mu_B^6)$ for the pressure and $\mathcal{O}(\mu_B^5)$ for $\rho_B$ computed on the lattice in Ref. \cite{Bazavov:2017dus}. 
%}
%\end{figure}
 %{\bf Presentation of the alternative methods to explore the QCD phase diagram at small chemical potentials, such as the Taylor expansion of the observables around $\mu_B/T=0$ and the analytical continuation of results obtained at imaginary chemical potential. Taylor expansion coefficients are now available up to order $(\mu_B/T)^6$. Thermodynamic quantities will be presented and their relevance for the experimental program at RHIC will be discussed. Cite the high density results of Kurkela.}

\subsection{QCD phase diagram}
Lattice QCD simulations predict that the phase transition from hadronic degrees of freedom to the QGP is a broad analytical crossover,  which happens gradually in the temperature range $T\simeq 145-165$ MeV \cite{Aoki:2006we,Aoki:2009sc,Borsanyi:2010bp,Bhattacharya:2014ara}; due to the nature of the transition, it is not possible to unambiguously define a transition temperature, and different observables yield slightly different values. The inflection point of the chiral condensate and the peak of the renormalized chiral susceptibility or
the disconnected chiral susceptibility yield a transition temperature $T=154\pm9$ \cite{Bazavov:2011nk}. The chiral crossover
temperature is closely related to the phase transition temperature in the limit of vanishing $u$ and $d$ quark masses \cite{Bazavov:2011nk,Ejiri:2009ac}. The deconfinement transition is well defined in the pure gauge limit, the renormalized Polyakov loop being the order parameter in that case. For physical values of the quark masses, the Polyakov loop still shows a rapid rise which leads to the identification of an inflection point, which is however renormalization-scheme dependent. It was recently pointed out that the negative derivative of the static quark free energy (which is defined as the logarithm of the Polyakov loop times the temperature)
has a peak at a temperature which coincides with the chiral transition temperature within errors \cite{Bazavov:2016uvm}.

Through a Taylor expansion or analytical continuation from imaginary chemical potential, it is possible to follow the behavior of the transition temperature at finite density: the temperature itself can be expanded in Taylor series as follows:
\begin{center}
\begin{equation}
~~~~~~~~~~~~~~~~~~\frac{T_c(\mu_B)}{T_c(0)}=1-\kappa\left(\frac{\mu_B}{T_c(0)}\right)^2+\lambda\left(\frac{\mu_B}{T_c(0)}\right)^4+...~~~.
\end{equation}
\end{center}
\begin{figure}[hb]
\begin{center}
\includegraphics[width=2.7in]{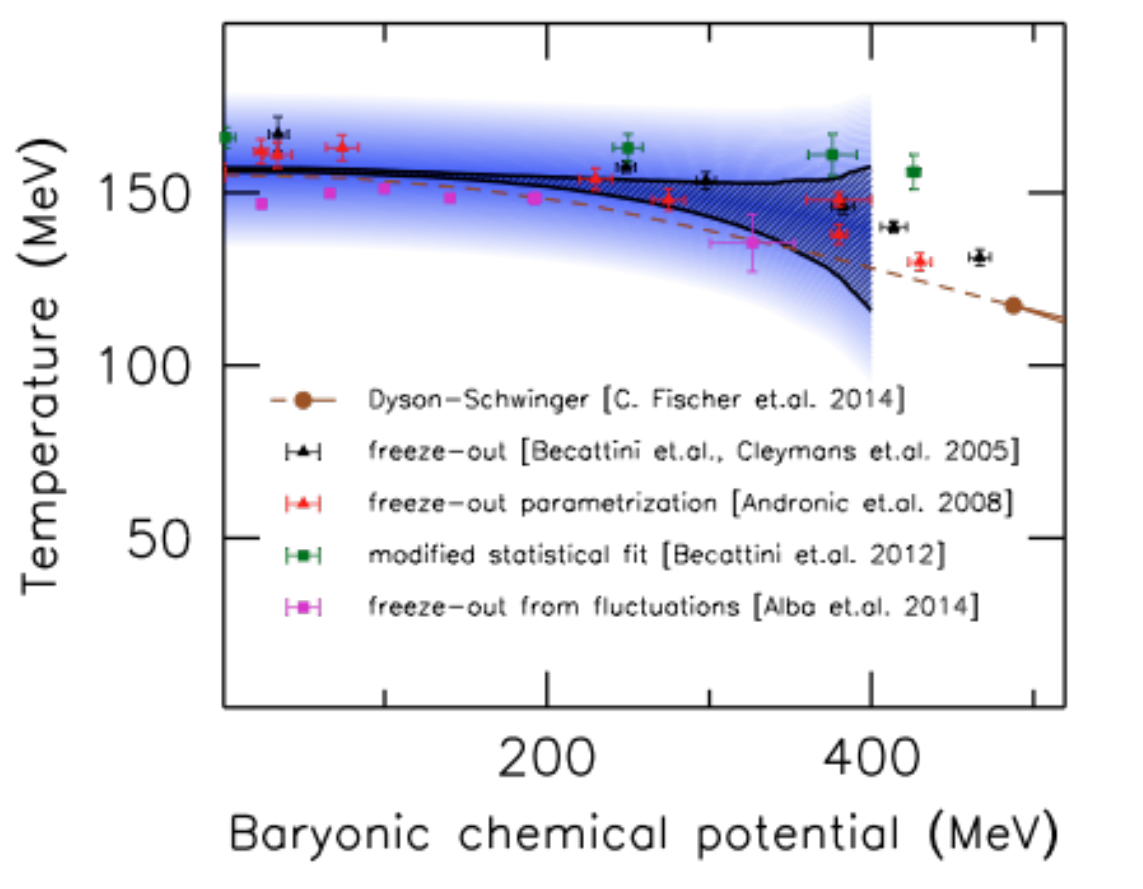}
\includegraphics[width=3.4in]{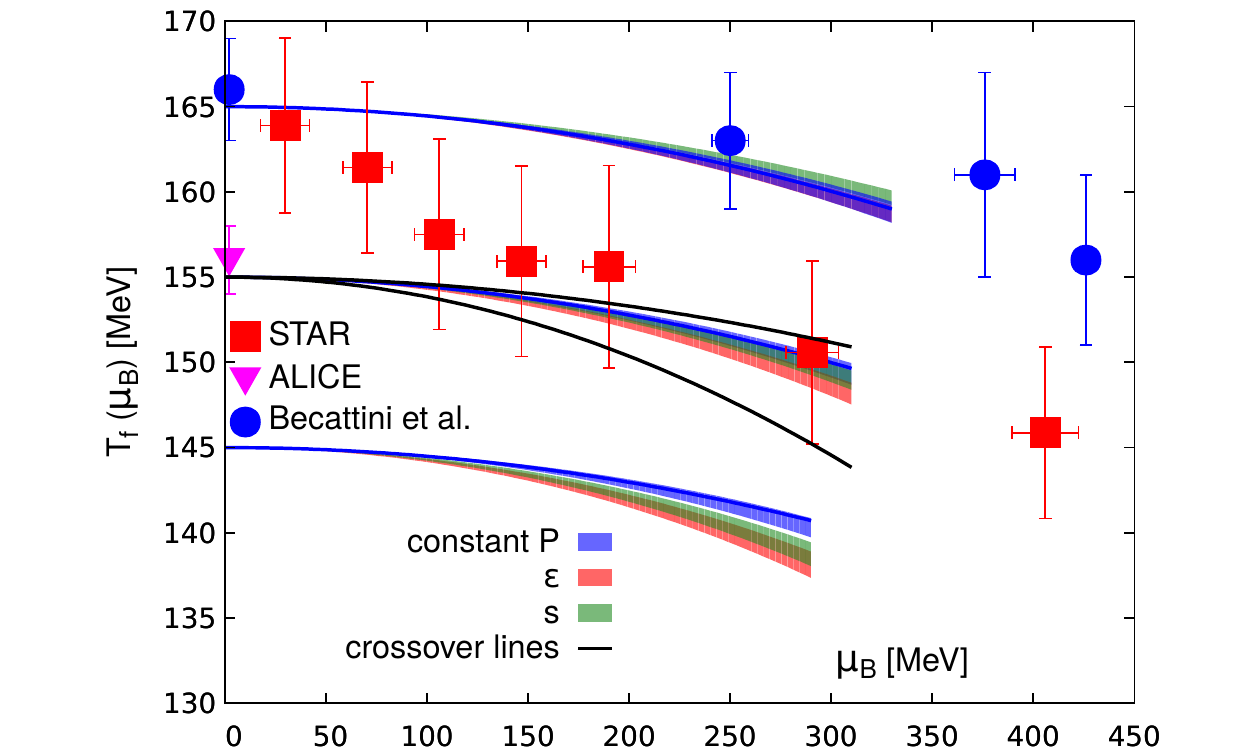}
\end{center}
%\vspace{-.5cm}
\caption{\label{fig6}
Left: The  phase  diagram  obtained in Ref. \cite{Bellwied:2015rza}.  The blue band indicates the width
of the transition.  The shaded black region shows the transition line obtained from the chiral
condensate.  The widening around 300 MeV is coming from the uncertainty of the curvature
and  from  the  contribution  of  higher  order  terms.  Also shown are some
selected non-lattice results: the Dyson-Schwinger result of Ref. \cite{Matias:2014jua} and the freeze-out data
of Refs. \cite{Cleymans:2004pp,Becattini:2005xt,Andronic:2008gu,Alba:2014eba}.
Right: From Ref. \cite{Bazavov:2017dus}. Lines of constant pressure,
energy density and entropy density versus temperature in (2+1)-flavor
QCD for three different initial sets of values fixed at $\mu_B=0$ 
and $T_0=145$~MeV, $155$~MeV and $165$~MeV, respectively. 
Data points show freeze-out temperatures determined by the
STAR Collaboration in the BES at RHIC (squares) \cite{Das:2014qca}
and the ALICE Collaboration at the LHC (triangle)
\cite{Floris:2014pta}. The circles denote hadronization temperatures
obtained by comparing experimental data on particle yields with a
hadronization model calculation \cite{Becattini:2016xct}. 
Also shown are two lines representing the current spread in determinations
of the $\mu_B$-dependence of the QCD crossover transition line.
}
\end{figure}
The curvature of the phase diagram, $\kappa$, turns out to be very small at $\mu_B=0$. Several results exist in the literature, which slightly differ from each other due mainly to the choice on how to treat the strange chemical potential. In the case in which the quark chemical potentials are all equal ($\mu_s=\mu_{u,d}$), and for $m_s/m_{u,d}=20$, the curvature obtained from the disconnected part of the renormalized susceptibility of the light quark chiral condensate is $\kappa=0.020(4)$ \cite{Cea:2015cya}. Using the chiral susceptibility and two different ways to extract the inflection point of the chiral condensate, the authors of Ref. \cite{Bonati:2015bha} found $\kappa=0.0135(20)$ at $\mu_S=0$. In Ref. \cite{Bellwied:2015rza}, a value of $\kappa=0.0149(21)$ was found from the chiral condensate, chiral susceptibility and strange quark susceptibility, in the case of strangeness neutrality ($\langle n_S\rangle=0$). The phase diagram obtained in Ref. \cite{Bellwied:2015rza} is shown in the left panel of Fig. \ref{fig6}. These results are in agreement with recent estimates of the curvature of lines of constant pressure, energy density and entropy density in the QCD phase diagram \cite{Bazavov:2017dus}: the latter are shown in the right panel of Fig. \ref{fig6}.

\begin{figure}[h]
\begin{center}
\includegraphics[width=3in]{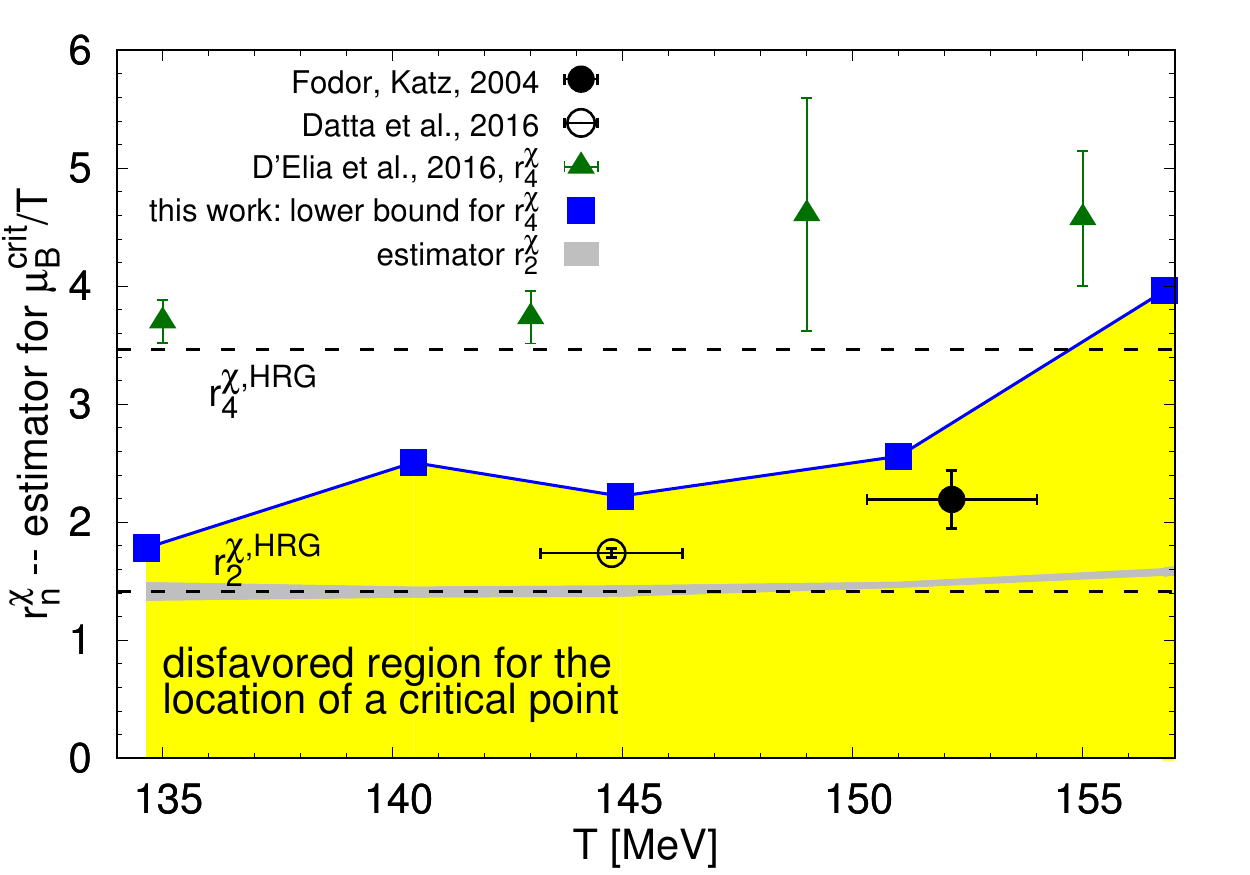}
\end{center}
%\vspace{-.5cm}
\caption{\label{fig7}
From Ref. \cite{Bazavov:2017dus}. Estimators for the radius of convergence of the Taylor series for net baryon-number fluctuations, $\chi^B_2(T; \mu_B)$, in the
case of vanishing electric charge and strangeness chemical potentials obtained on lattices with $N_t= 8$. Shown
are lower bounds for $r^\chi_{4,6}$ from Ref. \cite{Bazavov:2017dus} (squares) and from calculations
with an imaginary chemical potential (triangles) \cite{DElia:2016jqh}. The gray band shows an estimator from $r^{\chi}_{2,4}$. The figure also shows an estimate for the location of the critical point obtained from calculations using Taylor expansion (empty circle) \cite{Datta:2016ukp} and unimproved staggered fermions with a reweighting technique (full circle) \cite{Fodor:2004nz}. The two dashed lines are estimates based on $r^{\chi}_{4,6}$ and $r^{\chi}_{2,4}$ from the HRG model.}
\end{figure}

The coefficients of the Taylor series can provide information on the location of the QCD critical point \cite{Gavai:2004sd}. Since thermodynamic variables are non-analytic at the critical point, the radius of convergence of the Taylor series corresponds to the distance
from the series origin to the nearest non-analytic point.  If this point is on the
real $\mu$ axis, one has effectively located the CEP. The radius of convergence can be obtained from ratios of subsequent expansion coefficients for the free energy ($r^f_{n,m}$) or its derivatives with respect to $\mu_B$ ($r^{\chi}_{n,m}$) \cite{Karsch:2011yq}:
\begin{equation}
r^f_{n,m}(T)=\left|\frac{\frac{m!}{n!}\chi_n(T)}{\chi_m(T)}\right|^{1/(m-n)}~~~~~~~~~~~~~~~~r^{\chi}_{n,m}(T)=\left|\frac{\frac{(m-2)!}{(n-2)!}\chi_n(T)}{\chi_m(T)}\right|^{1/(m-n)}.
\end{equation}
These estimates necessarily coincide when $n$ and/or $m$ go to infinity, giving the radius of convergence of the series for a given $T$. The idea of using the Taylor expansion as a tool to locate the critical endpoint has been subject of discussion in the literature: while its validity has been tested e.g. in the 3D-Ising model \cite{York:2011km}, generally a successful extrapolation of the critical point requires high orders
in the cumulant expansion, as well as excellent control of errors. It is not clear whether the number of coefficients known in practice is enough to reliably constrain the location of the critical point. However, the expectation is that a consistent determination of the critical point appears when the estimators above agree with each other or show some sign of convergence. An example based on $r^\chi_{4,6}$ is shown in Fig. \ref{fig7} (from Ref. \cite{Bazavov:2017dus}), in which the two most recent sets of estimates for the radius of convergence from Refs. \cite{Bazavov:2017dus} and \cite{DElia:2016jqh} are shown. They exclude the presence of a critical point in the QCD phase diagram for $\mu_B/T\leq2$, in the temperature range 135 MeV $\leq T\leq$155 MeV.

Results for the QCD phase diagram in the presence of a magnetic field have been obtained e.g. in Refs. \cite{DElia:2010abb,Bali:2011qj}; the background magnetic field reduces the transition temperature; more recently, it has been pointed out that the transition becomes first order in the presence of asymptotically large magnetic fields \cite{Endrodi:2015oba}.

%{\bf Summary of our understanding of the QCD phase diagram, including value of the transition temperature and curvature of the transition line. Mention of ways to locate the critical point and state-of-the-art constraints on its existence/position.}
\subsection{Fluctuations of conserved charges}
Fluctuations of conserved charges are defined as
\begin{equation}
\chi_{lmn}^{BSQ}=\frac{\partial^{l+m+n}(p/T^4)}{(\partial\mu_B/T)^l(\partial\mu_S/T)^m(\partial\mu_Q/T)^n}.
\end{equation}
They can be calculated on the lattice as combinations of quark flavor fluctuations, through the following relationship between chemical potentials:
\begin{eqnarray}
\mu_u&=&\frac13\mu_B+\frac23\mu_Q
\nonumber\\
\mu_d&=&\frac13\mu_B-\frac13\mu_Q
\nonumber\\
\mu_s&=&\frac13\mu_B-\frac13\mu_Q-\mu_S.
\end{eqnarray}
The relevance of fluctuations for the physics of heavy ion collisions has been increasing in recent years. The higher order fluctuations of conserved charges are expected to diverge at the critical point, and therefore they have been proposed long ago as one of its possible experimental signatures \cite{Gavai:2008zr,Stephanov:1999zu,Cheng:2007jq}. For this reason, fluctuations have became one of the central measurements for the Beam Energy Scan at RHIC. Renewed interest in these observables has been stimulated also at small chemical potentials, due to the possibility of extracting freeze-out parameters of a heavy-ion collision from first principles, by comparing measurements to lattice QCD results \cite{Karsch:2012wm,Bazavov:2012vg,Borsanyi:2013hza} or of studying the chiral criticality through higher order fluctuations \cite{Redlich:2012xf}. Besides, linear combinations of fluctuations can be used to identify the effective degrees of freedom and study the chemical composition of the system at a given temperature and chemical potential \cite{Koch:2005vg,Bazavov:2013dta}.

The first continuum-extrapolated results for second order fluctuations of conserved charges at physical quark masses were presented in Ref. \cite{Borsanyi:2011sw} (almost-physical quark mass results are shown in Ref. \cite{Bazavov:2012jq}, heavier quark mass results are shown in Ref. \cite{Gavai:2005yk}) and later extended to selected fourth-order fluctuations and correlations \cite{Bellwied:2015lba,Ding:2015fca}. From these results it is evident that, at large temperatures, the observables are much closer to the ideal-gas limit, compared to the global thermodynamic observable presented in Section \ref{3.1}; also, these observables agree with perturbation theory predictions \cite{Haque:2014rua,Mogliacci:2013mca} for temperatures $T\geq 250$ MeV.

Before concentrating on the comparison of fluctuations of conserved charges with experiments, here we describe a couple of other possible applications for these observables. For example, it is possible to construct linear combinations of fluctuations which, in the low-temperature phase, select the contribution to thermodynamic quantities of hadrons according to their quantum numbers \cite{Bazavov:2013dta,Noronha-Hostler:2016rpd,Alba:2017mqu}. In the range of applicability of the HRG model, the pressure can be written as
\begin{eqnarray}
&&P_{tot}(T,{\mu}) = \sum _k P_k(T,\mu_k)
= \sum _k (-1)^{B_k+1} \frac{d_k T}{(2\pi) ^3} \times
\\ &&\int d^3\vec{p}
   \ln  \left(1+ (-1)^{B_k+1} \exp \left[ -\frac{(\sqrt{\vec{p} ^2+m_k^2}-\mu_k)}{T} \right] \right),
   \nonumber
\end{eqnarray}
where the index $k$ runs over all known baryons and mesons. Under the assumption that the Boltzmann approximation yields a good description of thermodynamic quantities, it is possible to re-write the pressure as
\begin{eqnarray}
\label{eq:pressure}
P(\hmu_B,\hmu_S) &=& P^{BS}_{00}+P^{BS}_{10}\cosh(\hmu_B)+P^{BS}_{01} \cosh(-\hmu_S) 
\nonumber \\
&+& P^{BS}_{11} \cosh(\hmu_B-\hmu_S)
+P^{BS}_{12} \cosh(\hmu_B-2\hmu_S)
\nonumber \\
&+& P^{BS}_{13} \cosh(\hmu_B-3\hmu_S) 
\;,
\end{eqnarray}
where $\hmu_i=\mu_i/T$, and the quantum numbers can be understood as absolute values.

The coefficients $P_{ij}^{BS}$ can be calculated either through simulations at imaginary chemical potential \cite{Alba:2017mqu} or through linear combinations of fluctuations that can be simulated directly at $\mu_B=0$ \cite{Bazavov:2013dta}. In the latter case, the definition of these quantities is not unique, and different combinations can be used, which agree in the hadronic phase but disagree as soon as strange quarks are liberated. This idea was used in Ref. \cite{Bazavov:2013dta} to find the onset of deconfinement for strange quarks.
\begin{figure}[h]
\begin{center}
\includegraphics[width=3in, height=2.6in]{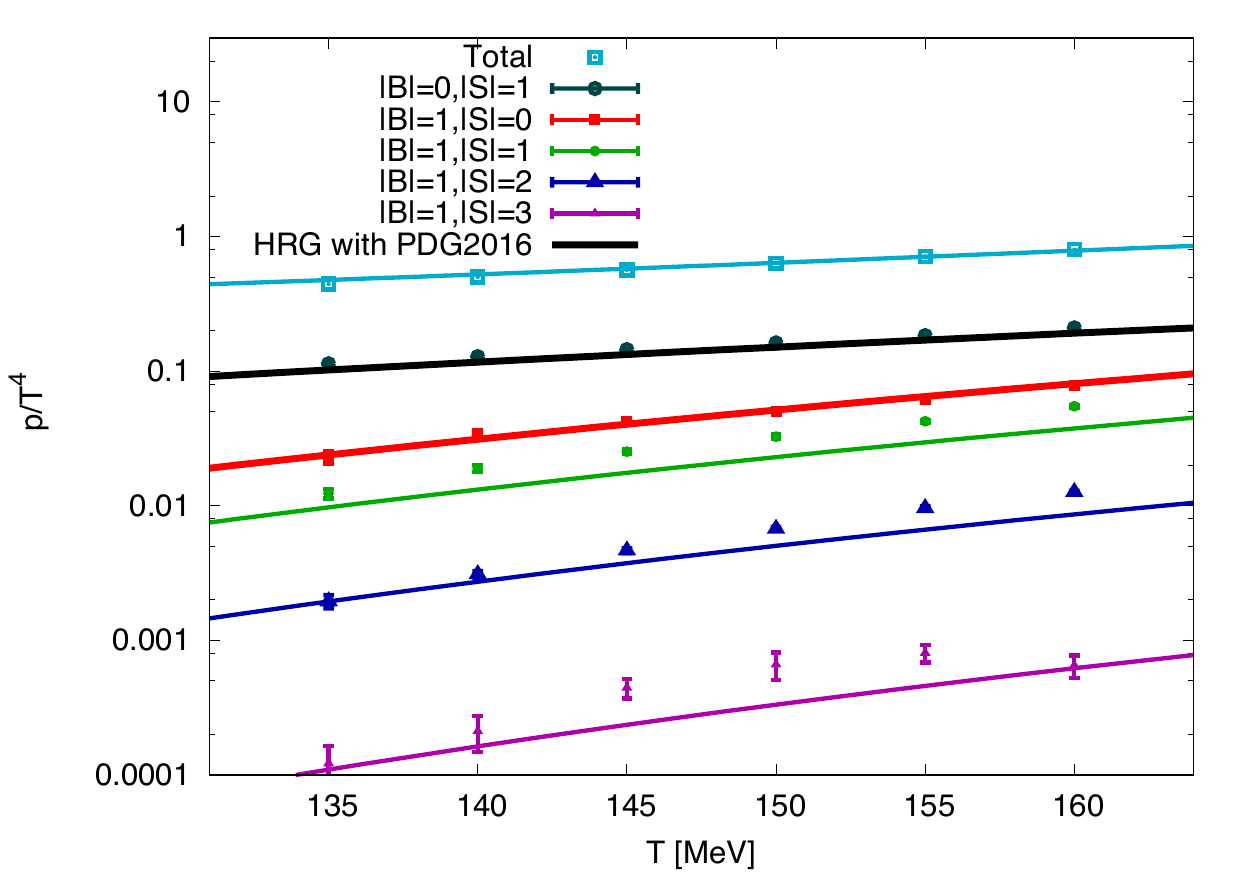}
\includegraphics[width=2.8in]{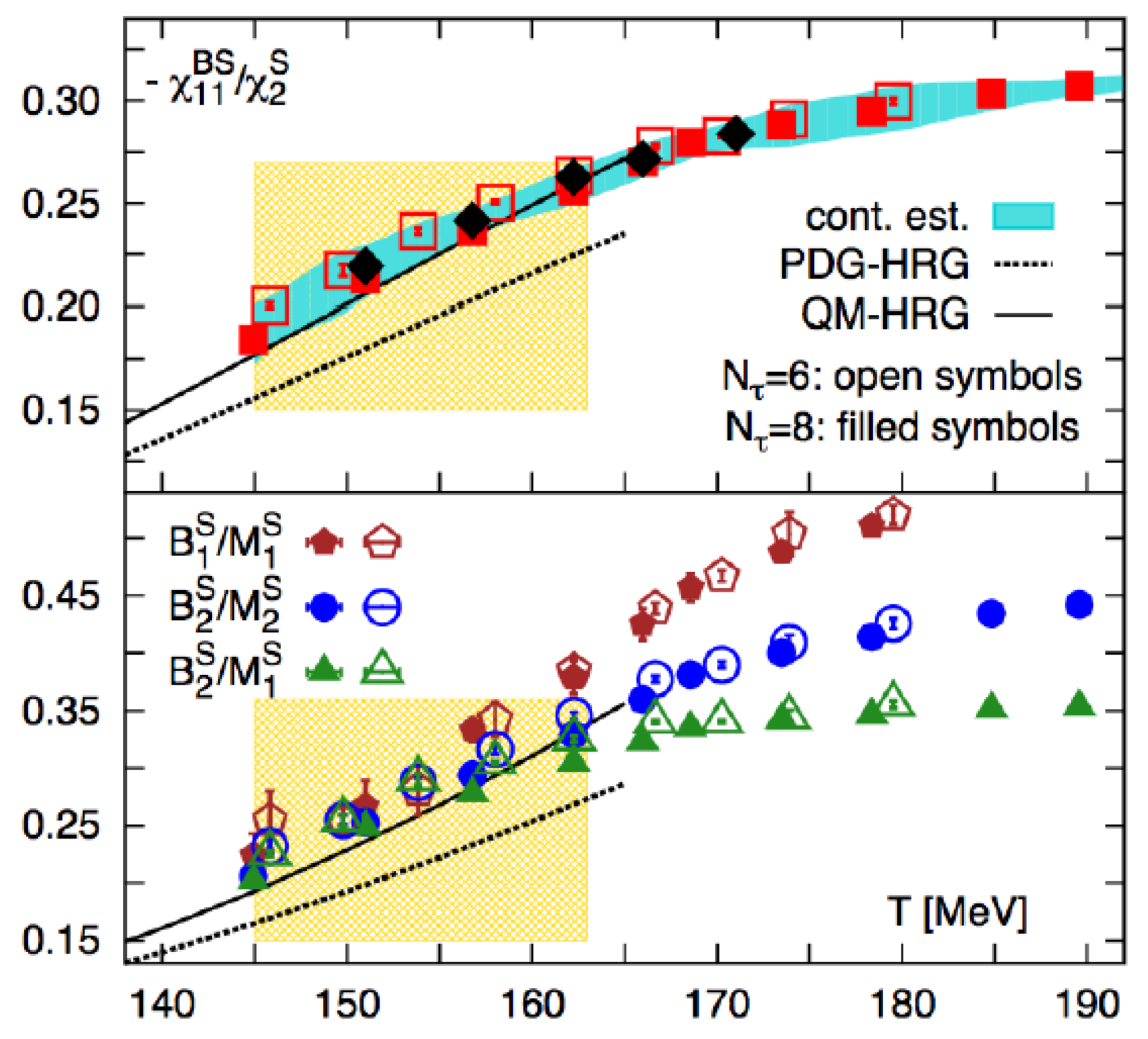}
\end{center}
%\vspace{-.5cm}
\caption{\label{fig8}
Left: From Ref. \cite{Alba:2017mqu}: Logarithmic plot illustrating the many
orders of magnitude covered by the values of the partial pressures. The total pressure is taken from Ref. \cite{Borsanyi:2013bia}.
In all cases, the solid lines correspond to the HRG model results based on the PDG2016 spectrum. Right: From Ref. \cite{Bazavov:2014xya}: $BS$ correlations normalized to the second cumulant of net strangeness fluctuations. Results are from (2+1)-flavor lattice QCD calculations. The band depicts the improved estimate for the
continuum result. The solid line is the result for this observable based on the PDG states, the dotted line includes states predicted by the Quark Model \cite{Capstick:1986bm,Ebert:2009ub}. 
}
\end{figure}

Extracting these coefficients through imaginary chemical potentials simulations considerably reduces the uncertainty of the lattice results, and allowed the authors of Ref. \cite{Alba:2017mqu} to extract very precise results for the pressures of the single hadronic families. These results are shown in the left panel of Fig. \ref{fig8}. This analysis has been stimulated by Ref. \cite{Bazavov:2014xya}, in which it was pointed out that the ratio of fluctuations $\chi_{11}^{BS}/\chi_2^S$ does not agree with the HRG model predictions already at low temperatures, indicating the need for more strange hadronic states with respect to the ones listed in the PDG (see the right panel of Fig. \ref{fig8}). In Ref. \cite{Alba:2017mqu}, it was possible to identify the flavor content of the missing hadrons, which will be useful to guide their future experimental searches.

The most interesting feature of fluctuations is that they can be related to the moments of the distribution of the corresponding conserved charges (mean $M$, variance $\sigma^2$, skewness $S$ and kurtosis $\kappa$) through the following formulas
\begin{eqnarray}
M=\chi_1~~~~&&~~~~\sigma^2=\chi_2
\nonumber\\
S=\chi_3/\chi_2^{3/2}~~~~&&~~~~\kappa=\chi_4/\chi_2^2.
\end{eqnarray}
Usually ratios are defined, so that the volume factor cancels out in the theoretical definition of fluctuations, and they become functions only of $T$ and $\mu_B$:
\begin{eqnarray}
M/\sigma^2=\chi_1/\chi_2~~~~~&&~~~~~~S\sigma=\chi_3/\chi_2
\nonumber\\
S\sigma^3/M=\chi_3/\chi_1~~~~~&&~~~~~~\kappa\sigma^2=\chi_4/\chi_2.
\end{eqnarray}

The idea is that the particle multiplicity and their fluctuations are fixed at the chemical freeze-out, so that their experimental value corresponds to the temperature and chemical potential of that particular moment in the evolution of the system. An interesting question is whether the fluctuation observables in a heavy-ion collision will reflect a sharp freeze-out, or whether the latter happens over a broader, possibly flavor-dependent, temperature range. This idea can be tested by extracting the freeze-out parameters from the fluctuations of different conserved charges, and comparing the results.
The idea of fluctuations of a conserved charge seems to be a contradiction in itself; indeed, if we were able to cover the entire solid angle with the experimental apparatus, we would measure the same amount of net-baryons, net-electric charge and net-strangeness which were there at the moment of the collision. However, by studying a sufficiently small subsystem, conserved quantities can fluctuate: the small system can exchange conserved charges with the rest of the system. This is similar to the assumptions which govern a thermal system in the Grand-Canonical Ensemble, which corresponds to lattice QCD calculations \cite{Koch:2008ia}. A discussion of this and other conditions, for a meaningful comparison with experimental measurements, will be presented in the next subsection.

The quantities that are utilized, in order to extract the freeze-out temperature
and baryon chemical potential, are ratios of fluctuations of conserved charges, such as $\chi_3^Q/\chi_{1}^Q$,
$\chi_1^Q/\chi_{2}^Q$, $\chi_3^B/\chi_{1}^B$,
$\chi_1^B/\chi_{2}^B$ etc. These are calculated at some $(\mu_B,\mu_Q,\mu_S)$ point, which is defined
by the pyhsical conditions which match the experimental situation, namely which satisfy $\langle n_S\rangle=0$ and $\langle n_Q\rangle=0.4\langle n_B\rangle$. The first terms of their Taylor expansion around $\mu_B=0$ read:
\begin{eqnarray}
\hspace{-2cm}
&&R_{31}^Q(T,\mu_B)=\frac{\chi_3^Q(T,\mu_B)}{\chi_1^Q(T,\mu_B)}=
\frac{\chi_{31}^{QB}(T,0)+\chi_{4}^{Q}(T,0)q_1(T)+\chi_{31}^{QS}(T,0)s_1(T)}{\chi_{11}^{QB}(T,0)+\chi_{2}^{Q}(T,0)q_1(T)+\chi_{11}^{QS}(T,0)s_1(T)}+\mathcal{O}(\mu_B^2)\nonumber\\
\hspace{-2cm}
&&R_{12}^Q(T,\mu_B)=\frac{\chi_1^Q(T,\mu_B)}{\chi_2^Q(T,\mu_B)}=
\frac{\chi_{11}^{QB}(T,0)+\chi_{2}^{Q}(T,0)q_1(T)+\chi_{11}^{QS}(T,0)s_1(T)}{\chi_2^Q(T,0)}\frac{\mu_B}{T}+\mathcal{O}(\mu_B^3).
\nonumber\\
\hspace{-2cm}
&&
\end{eqnarray}
The leading order in $\chi_3^Q/\chi_1^Q$ is independent of $\mu_B$, which makes $R_{31}^Q$ the ideal quantity to extract the freeze-out temperature. Once $T_f$ has been obtained with this method, the ratio $R_{12}^Q$ can then be used to determine $\mu_B$.
This, and other possible methods to extract the freeze-out parameters from first principles, will be discussed in the following subsection.

%{\bf The fluctuations of conserved charges are introduced. Their relevance is explained, both as tools to identify the chemical composition of the system and as experimentally measurable observables to extract the freeze-out parameters.}
\subsubsection{Results from lattice simulations and comparison to experiment~~~~~~~~~~~~~~~~~~~~~~~~}

The comparison of lattice QCD results and experimental measurements for fluctuations of conserved charges is an exciting development which was not possible until recent years, due to the fact that simulations were performed on coarse lattices and for unphysical values of the parameters. Nevertheless, a direct link between a complex dynamical system and a Grand Canonical Ensemble in thermal equilibrium represents an ambitious goal which needs the fulfillment of several conditions in order to be meaningful.

As already mentioned, the rapidity cuts introduced by the detector can turn a Canonical into a Grand Canonical Ensemble, provided that the window has the right size, namely this subsystem should be small enough so that total charge conservation does not suppress the signal, but large enough to be still suitably described as a thermodynamic system in equilibrium:
\begin{equation}
\Delta Y_{tot}\gg\Delta Y_{accept}\gg\Delta Y_{kick}~~~~~~~~~~~~~\Delta Y_{accept}\gg\Delta Y_{corr}
\end{equation}
where $\Delta Y_{tot}$ is the range for the total charge multiplicity distribution, $\Delta Y_{accept}$ is the experimental acceptance window, $\Delta Y_{kick}$ is the typical rapidity shift that charges receive during and after hadronization and $\Delta Y_{corr}$ is the charge correlation length characteristic of the physics of interest. The first criterion ensures that total charge conservation does not suppress the signal and that the signal survives hadronization and the hadronic phase, while the second one is necessary in order to be sensitive to the relevant physics \cite{Koch:2008ia}. Currently, the maximum rapidity window used in experimental publications of fluctuations is $|\Delta y|\leq0.5$. Recently, first preliminary studies of the rapidity-dependence of the fluctuation measurements have become available. Hopefully, thanks to these studies, in the near future it will be possible to identify the ideal rapidity window which allows to treat the experimental system as a Grand Canonical Ensemble.

\begin{figure}[h]
\begin{center}
\includegraphics[width=2.8in]{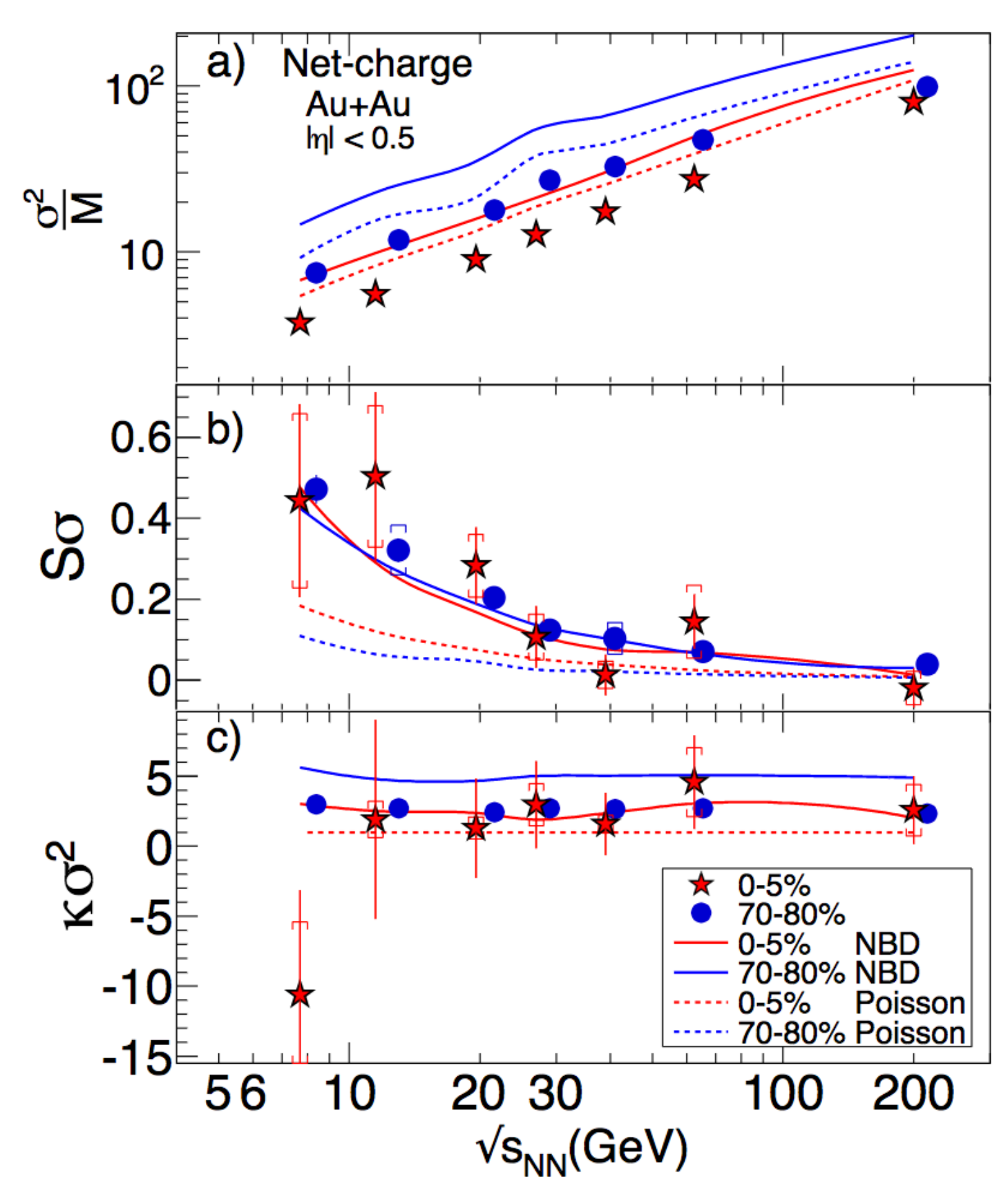}
\includegraphics[width=2.8in]{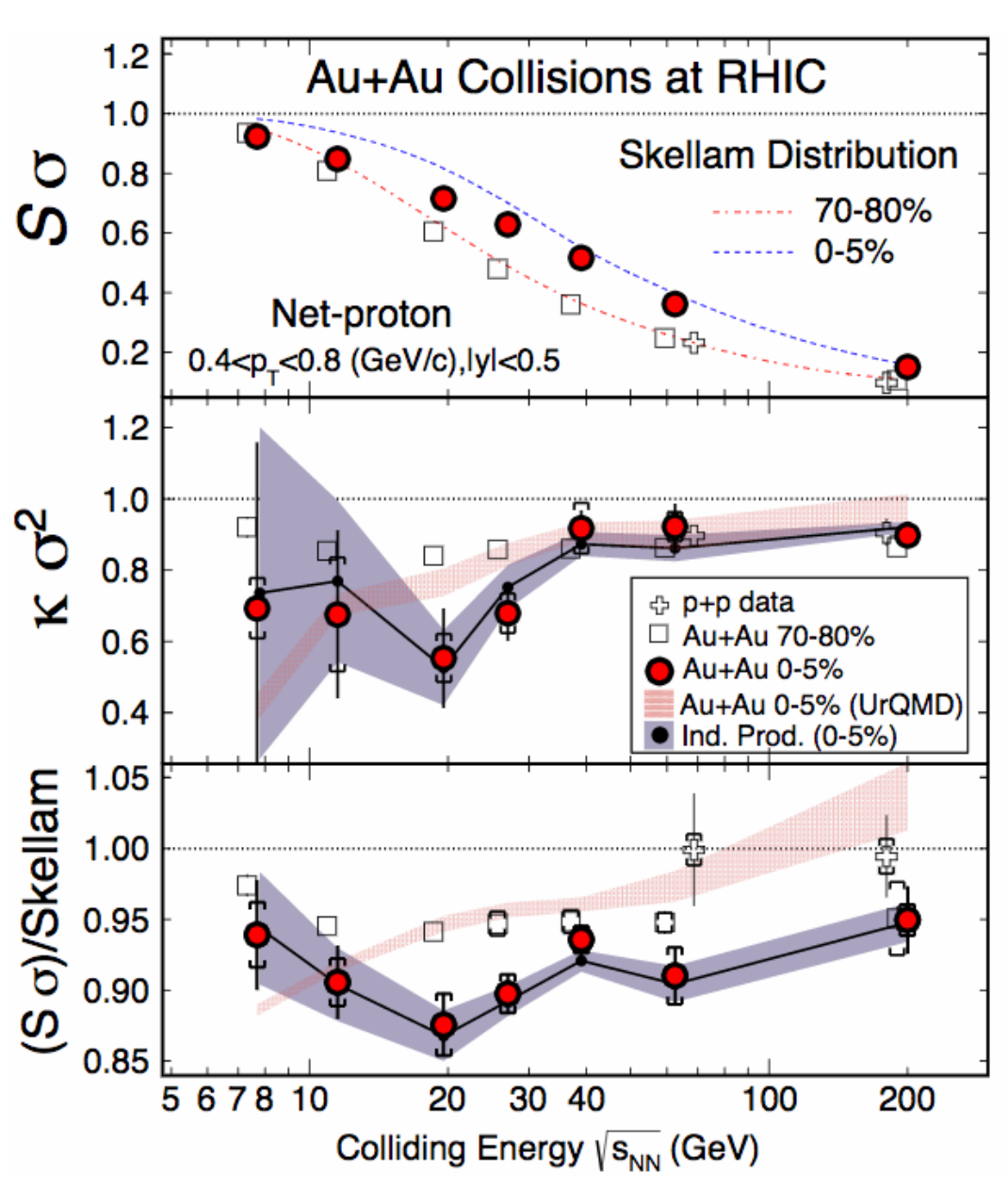}
\end{center}
%\vspace{-.5cm}
\caption{\label{fig9}
Left: Net-electric charge fluctuation ratios published by the STAR collaboration \cite{Adamczyk:2014fia}. Right: net-proton fluctuations published by the STAR collaboration \cite{Adamczyk:2013dal}.
}
\end{figure}

The transverse-momentum dependence of fluctuations has also been investigated by the STAR collaboration: in their original publication of proton-number fluctuations, they had chosen 0.4 GeV$\leq p_T\leq$ 0.8 GeV \cite{Adamczyk:2013dal}, while for net-electric charge the $p_T$ range was 0.2 GeV $\leq p_T\leq$ 2.0 GeV, after removing protons and anti-protons with $p_T<$ 400 MeV/c \cite{Adamczyk:2014fia}. The latter was done in order to remove spallation protons which come from interactions of the beam with the beam pipe. The published results for net-electric charge and net-proton fluctuations are shown in the two panels of Fig. \ref{fig9}. More recently, net-proton fluctuations have been presented, with a momentum cut 0.4 GeV$\leq p_T\leq$ 2.0 GeV \cite{Luo:2015ewa}, which have generated quite some excitement in the community, since the higher-order fluctuations show a non-monotonic behavior which might be a signature of criticality \cite{Stephanov:2011pb,Ling:2015yau}.

Possible  experimental
sources of non-thermal fluctuations are corrected for in
the STAR data analysis:  the centrality-bin-width correction  method  minimizes  effects  due  to  volume  variation
because of finite centrality bin width (such effects have been studied e.g. in \cite{Skokov:2012ds,Braun-Munzinger:2016yjz,Begun:2017sgs});  the moments are
corrected for the finite reconstruction efficiency based on
binomial probability distribution \cite{Bzdak:2012ab}. A $p_T$-dependent efficiency correction method \cite{Bzdak:2013pha} has been recently implemented; multiplicity-dependent and non-binomial efficiency corrections have also been studied \cite{Bzdak:2016qdc}, as well as the effect of baryon number conservation on the cumulants of net-proton distribution \cite{Bzdak:2012an}.

Final-state  interactions in the hadronic phase and non-equilibrium effects might become relevant and affect fluctuations \cite{Becattini:2016xct,Rapp:2000gy,Rapp:2001bb,Rapp:2002fc,Steinheimer:2012rd,Becattini:2012xb,Steinheimer:2016cir}; a fundamental
check in favor of the equilibrium scenario is e.g. the consistency between the freeze-out parameters yielded by different quantum numbers, like  electric charge and baryon number.

One more caveat is in order, since experimentally only the net-proton multiplicity distribution is measured,  as
opposed  to  the  lattice  net-baryon  number  fluctuations. It was shown that, once the effects of resonance
feed-down and isospin randomization are taken into account  \cite{Kitazawa:2011wh,Kitazawa:2012at},  the  net-proton  and  net-baryon  number fluctuations are numerically very similar, at least in the case of low-order fluctuations \cite{Nahrgang:2014fza}.

The above discussion shows that a safe and meaningful comparison between experimental fluctuation measurements and lattice QCD results can be quite challenging and several effects need to be properly understood. Nevertheless, such comparisons have been performed and have led to quite interesting and consistent results, which will be reviewed in this subsection.

\begin{figure}[h]
\begin{center}
\includegraphics[width=3in]{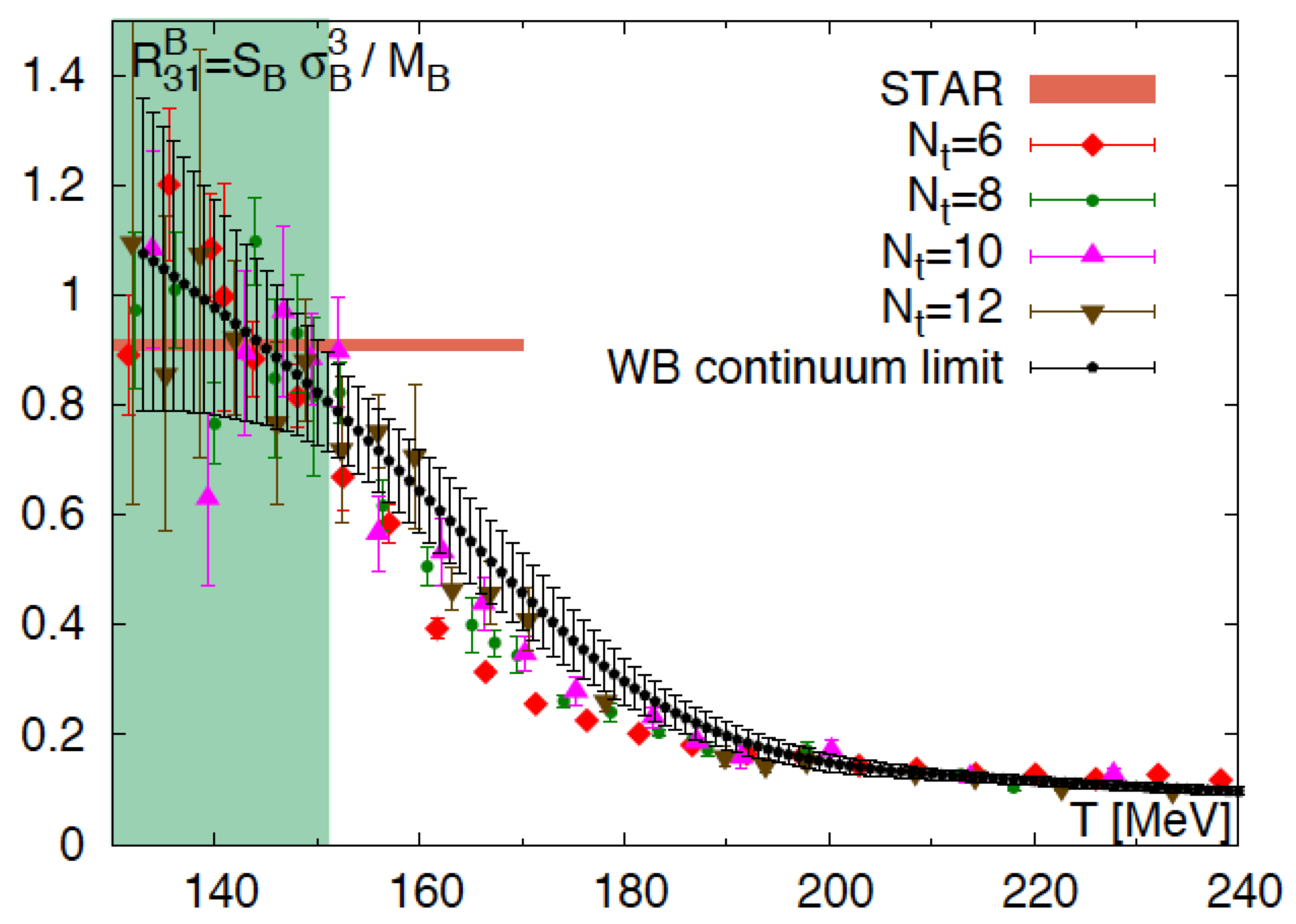}
\includegraphics[width=2.9in]{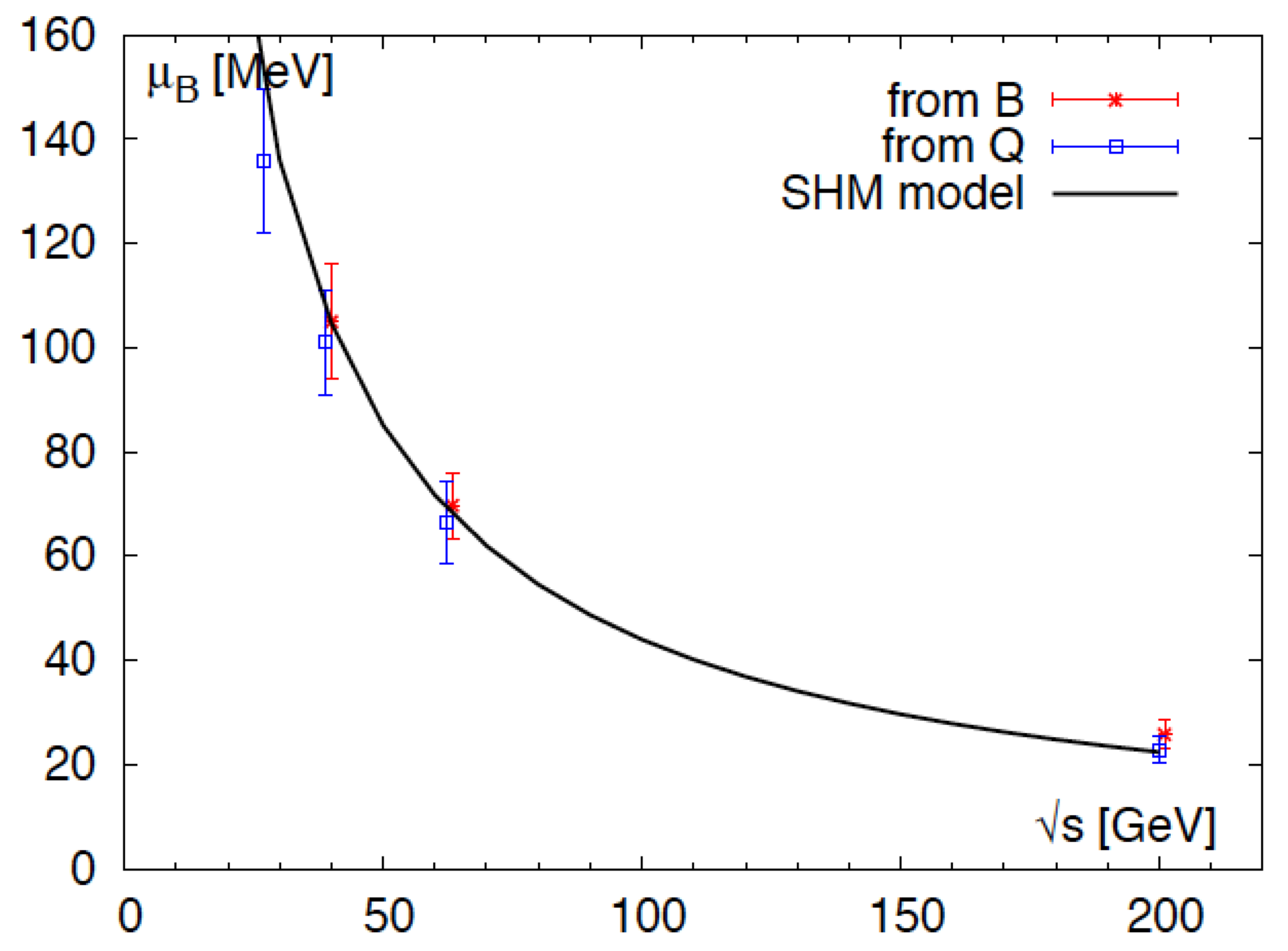}
\end{center}
%\vspace{-.5cm}
\caption{\label{fig10}
From Ref. \cite{Borsanyi:2014ewa}. Left: Baryonic $S\sigma^3/M=\chi_3/\chi_1$: the black dots correspond to the continuum extrapolated lattice QCD results from the WB collaboration \cite{Borsanyi:2014ewa}; the orange band is the STAR collaboration value, obtained by averaging the measurements corresponding to the four highest collision energies and for a centrality $0-10$\% from Ref. \cite{Adamczyk:2013dal}. Right: comparison between the freeze-out chemical potentials obtained from the baryon number (red stars) and electric charge (blue squares) $\chi_1/\chi_2$ fit. Also shown is the black curve corresponding to the Statistical Hadronization Model fit to particle yields \cite{Andronic:2008gu,Andronic:2005yp}.
}
\end{figure}

The first comparison between lattice QCD results and experimental fluctuations was performed in Refs. \cite{Bazavov:2012vg,Borsanyi:2013hza}, in which preliminary electric charge fluctuations from the STAR collaboration were used for the comparison. In Ref. \cite{Borsanyi:2014ewa} a separate analysis for electric charge and baryon number fluctuations was performed, which led to consistent values of the freeze-out conditions. This is a non-trivial test which seems to imply thermal equilibrium at the chemical freeze-out, validating the comparison with lattice QCD results. The chemical freeze-out temperature resulting from this analysis turned out to be $T_f\leq 151$ MeV. The left panel of Fig. \ref{fig10} shows the comparison of lattice results and experimental measurement for the baryonic $\chi_3/\chi_1$, from which the upper bound for $T_f$ was extracted. The right panel of Fig. \ref{fig10} shows the freeze-out chemical potentials from baryon number and electric charge fluctuation fits, as functions of the collision energy.

\begin{figure}[h]
\begin{center}
\includegraphics[width=3in,height=2.8 in]{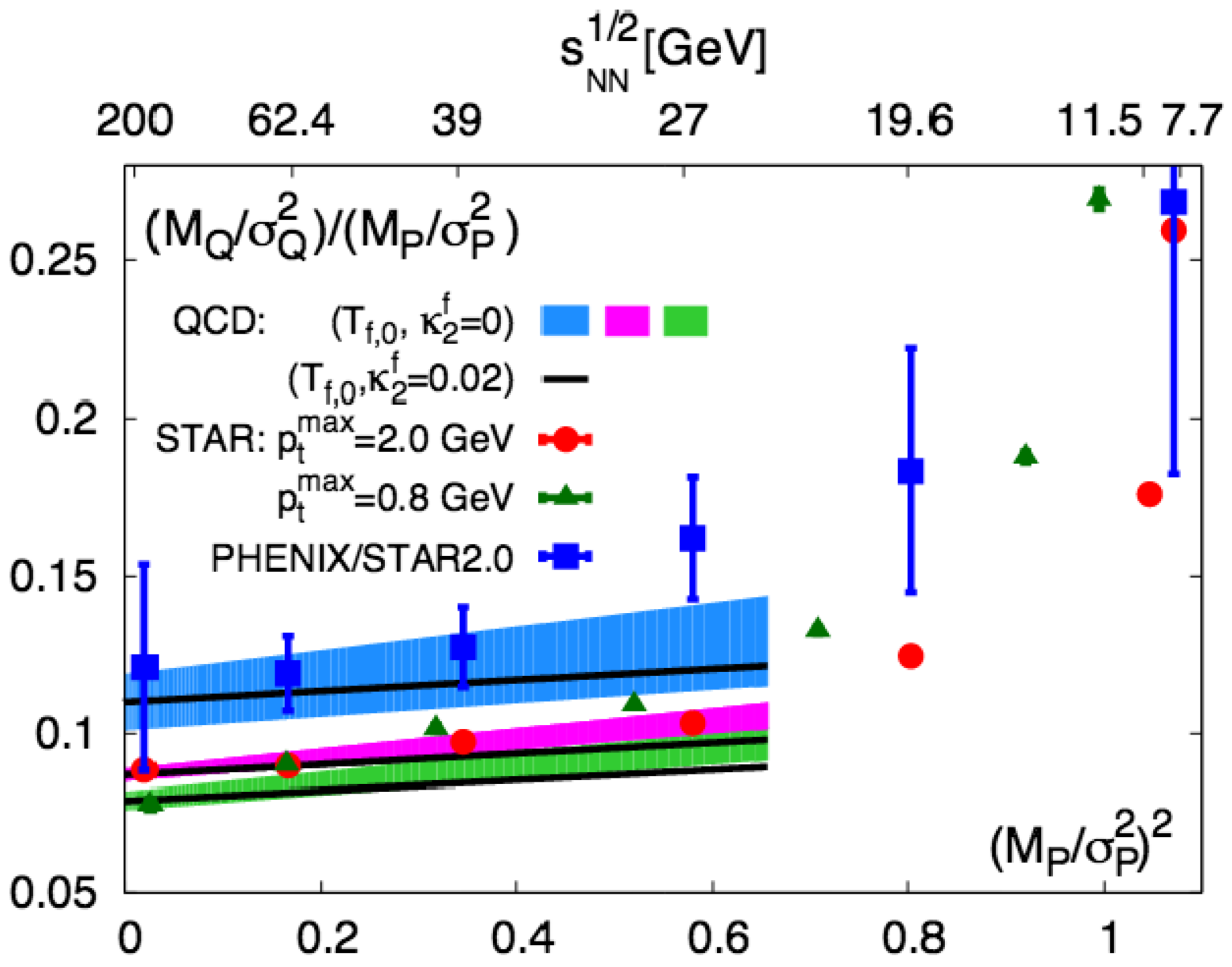}
\includegraphics[width=2.9in]{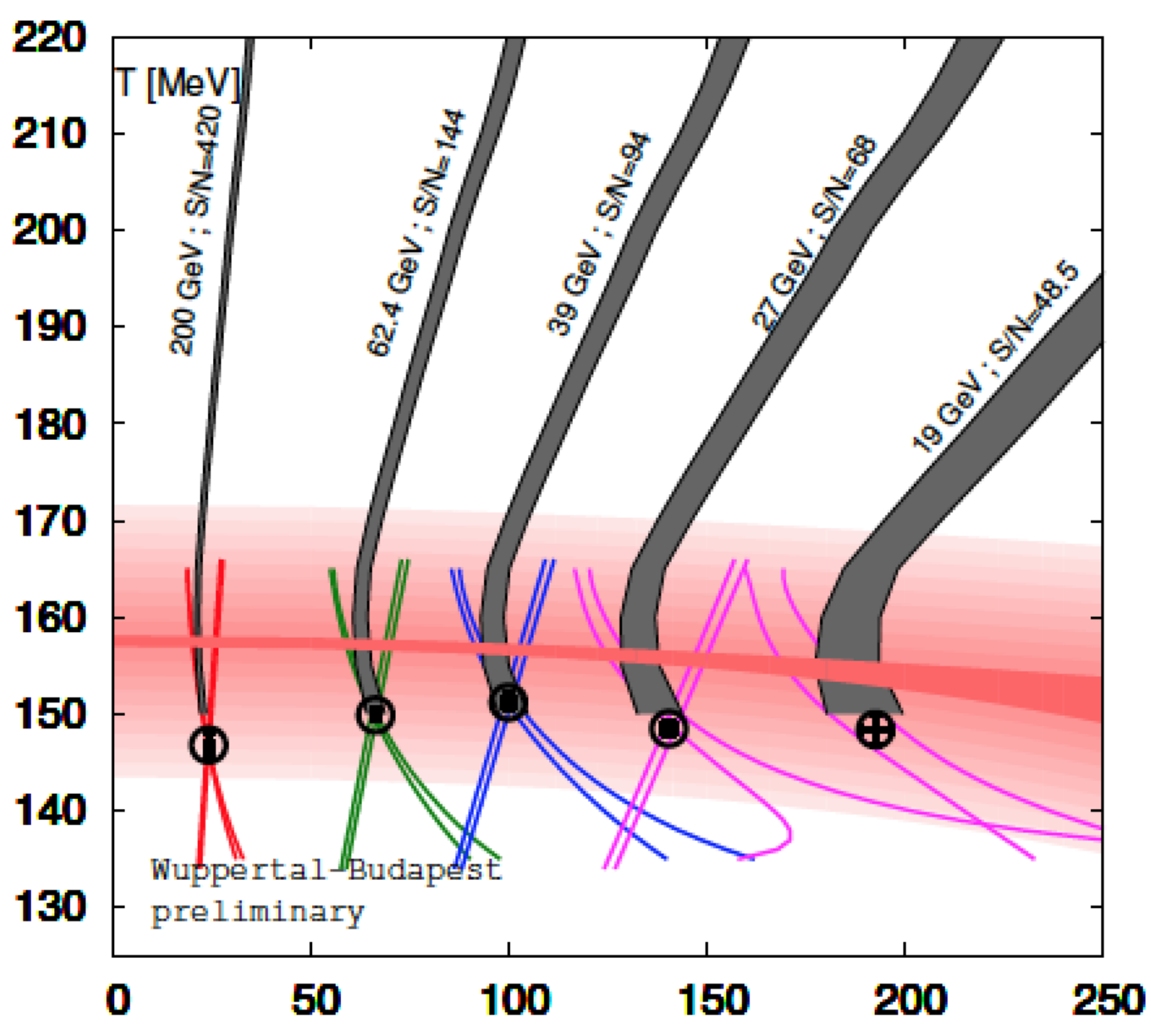}
\end{center}
%\vspace{-.5cm}
\caption{\label{fig11}
Left: From Ref. \cite{Bazavov:2015zja}: the ratio of ratios $\chi_1/\chi_2$ for net-electric charge and net-proton fluctuations measured by the STAR and PHENIX Collaborations \cite{Adamczyk:2013dal,Adamczyk:2014fia,Luo:2015ewa,Adare:2015aqk}. Right: Preliminary results of the WB collaboration \cite{Ratti:2016jgx}. The colored lines
are the contours at constant mean/variance ratios of the net electric charge from lattice simulations. The contours that correspond to
STAR data intersect in the freeze-out points of Ref. \cite{Alba:2014eba}. The red band is the QCD phase diagram shown in the left panel of Fig. \ref{fig6}. Also shown are the isentropic contours that match the chemical freeze-out data.
}
\end{figure}

More recently, the authors of Ref. \cite{Bazavov:2015zja} performed a fit to the ratio of ratios
of $\chi_1/\chi_2$ (mean/variance) for electric charge and proton number and were able to obtain both the freeze-out
temperature and the curvature of the freeze-out line. The left panel of Fig. \ref{fig11} shows the ratio of ratios
of $\chi_1/\chi_2$ for electric charge and proton number used for this fit. The value of the freeze-out temperature ($T_f = (147\pm2)$
MeV) is in agreement with the one obtained in Ref. \cite{Borsanyi:2014ewa}. The curvature value is found to be $\kappa_f<0.011$, compatible with lattice QCD results on the curvature of the QCD transition line.

The WB collaboration recently performed a combined fit of $\chi_1/\chi_2$ for electric charge and baryon number and found freeze-out temperature and chemical potential corresponding to the five highest collision energies at RHIC (from $\sqrt{s}=200$ GeV to 19.6 GeV) \cite{Ratti:2016jgx}. These results are in agreement with the freeze-out temperature and curvature values extracted in Ref. \cite{Bazavov:2015zja} and, remarkably, with the same fluctuation analysis performed in an HRG model in which the effects of resonance decay and regeneration, and the kinematic cuts corresponding to the experimental situation, were taken into account \cite{Alba:2014eba}.

The left panel of Fig. \ref{fig11} also shows data from the PHENIX collaboration \cite{Adare:2015aqk}, which published results only for net-charge fluctuations with 0.3 GeV$\leq p_T\leq$2.0 GeV and $|\eta|\leq0.35$; the discrepancy between these results and the STAR ones can be understood in terms of the different $p_T$ and rapidity windows \cite{Bazavov:2015zja}.

\begin{figure}[h]
\begin{center}
\includegraphics[width=6.2in]{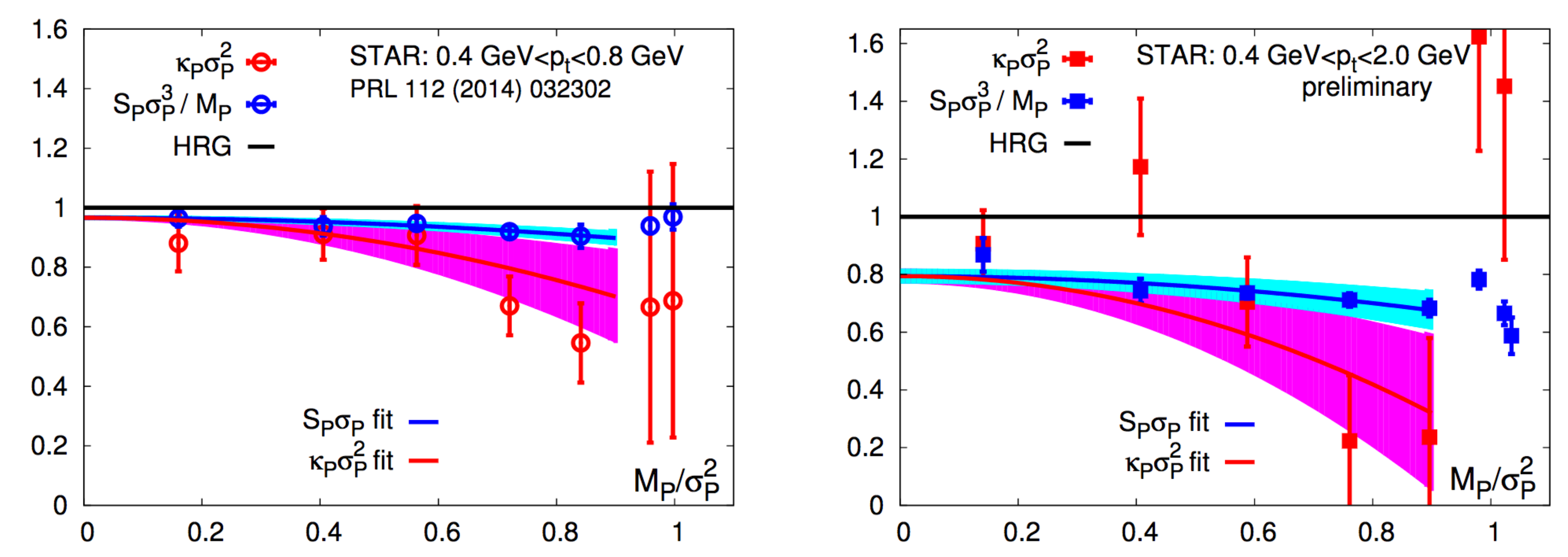}
\end{center}
%\vspace{-.5cm}
\caption{\label{fig12}
From Ref. \cite{Karsch:2016yzt}. Net-proton $S\sigma^3/M$ and $\kappa\sigma^2$ measured by STAR with 0.4GeV $\leq p_T\leq$0.8 GeV (left) and 0.4 GeV$\leq p_T\leq$2.0 GeV (right) plotted against $M_p/\sigma_p^2$. Curves are fits to the data, constrained by the condition $S\sigma^3/M=\kappa\sigma^2$ at $M_p/\sigma_p^2=0$.}
\end{figure}

Most results discussed so far concerned lower-order fluctuations and their ratios. They are more suitable to extract the freeze-out parameters because they are measured more precisely and are less affected by kinematic cuts and other effects. Besides, extrapolating the kurtosis to finite chemical potential at leading order requires the knowledge of the sixth-order coefficients in the Taylor expansion of the pressure. The first results on higher order fluctuations at finite chemical potential are becoming available. In Ref. \cite{Karsch:2016yzt} it was shown that the trend shown by both sets of STAR data for $\kappa\sigma^2$ can be understood by expanding these observables in powers of $\mu_B/T$ up to $\mathcal{O}(\mu_B^2)$.
Such expansions for $S\sigma^3/M$ and $\kappa\sigma^2$ read:
\begin{eqnarray}
\frac{S_B\sigma_B^3}{M_B}&=&\frac{\chi_4^B+s_1\chi_{31}^{BS}+q_1\chi_{31}^{BQ}}{\chi_2^B+s_1\chi_{11}^{BS}+q_1\chi_{11}^{BQ}}=r_{31}^{B,0}+r_{31}^{B,2}\left(\frac{\mu_B}{T}\right)^2+\mathcal{O}(\mu_B^4)
\nonumber\\
\kappa_B\sigma_B^2&=&\frac{\chi_4^B(T,0)}{\chi_2^B(T,0)}+\mathcal{O}(\mu_B^2)=r_{42}^{B,0}+r_{42}^{B,2}\left(\frac{\mu_B}{T}\right)^2+\mathcal{O}(\mu_B^4).
\end{eqnarray}

Results for $r_{31}^{B,2}$ and $r_{42}^{B,2}$ as functions of the temperature have been calculated in Ref. \cite{Karsch:2016yzt}; in spite of the relatively large uncertainty on the results, it is clear that these coefficients are negative at the temperature around the transition, and that  $r_{42}^{B,2}$ is three times larger than $r_{31}^{B,2}$, which explains why $\kappa_B\sigma_B^2$ drops faster than $S\sigma^3/M$ as the chemical potential is increased. This is indeed the trend manifested by the STAR data, which is evident in the two panels of Fig. \ref{fig12} that show the two $p_T$ cuts implemented by STAR.

\begin{figure}[h]
\begin{center}
\includegraphics[width=3in,height=2.6 in]{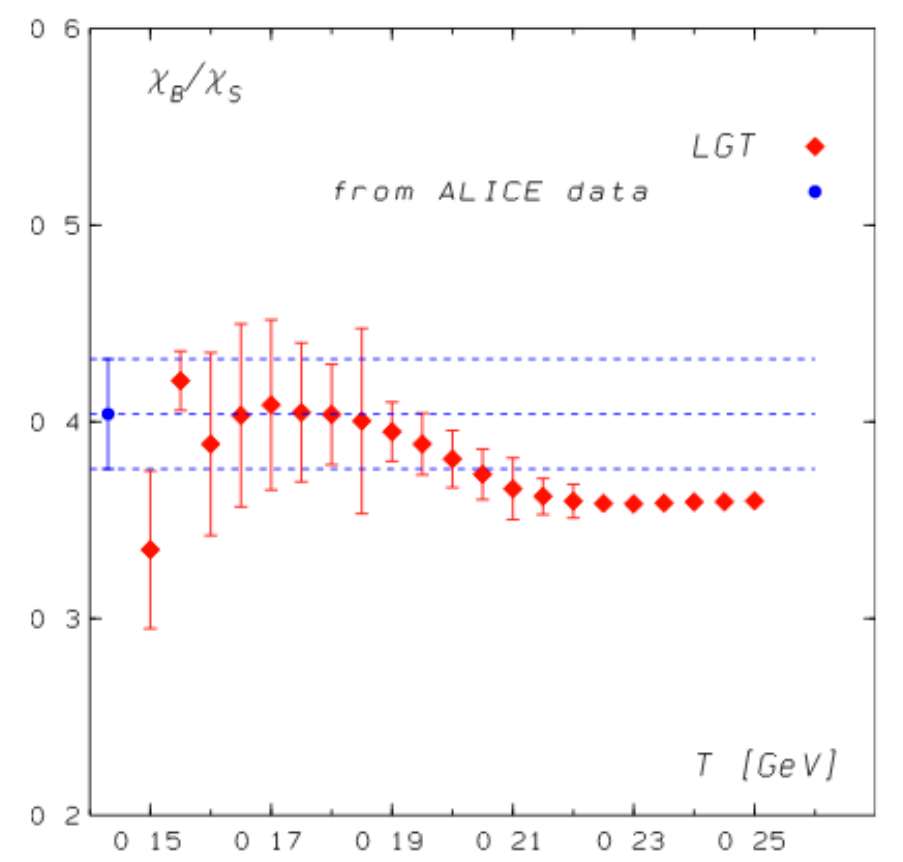}
\includegraphics[width=3.1in, height=2.6in]{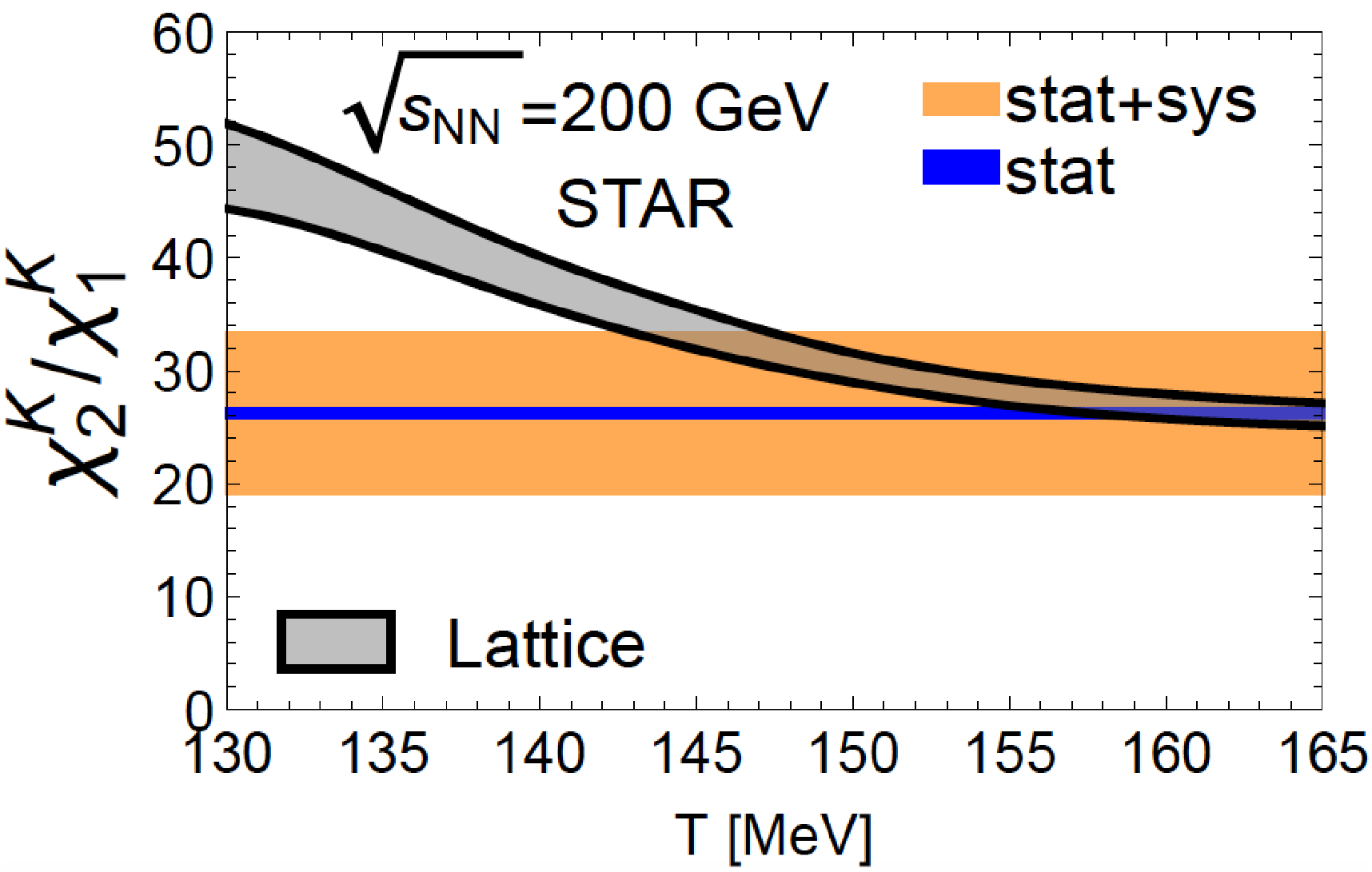}
\end{center}
%\vspace{-.5cm}
\caption{\label{fig13}
Left: From Ref. \cite{Braun-Munzinger:2014lba}: the ratio $\chi_2^B/\chi_2^S$: comparison between ALICE measurement (band) and lattice QCD simulation results from Ref. \cite{Bazavov:2012jq} (red dots). Right: From Ref. \cite{Noronha-Hostler:2016rpd}: $\chi_2/\chi_1$ for charged Kaons: comparison between the preliminary STAR measurement at $\sqrt{s}=200$ GeV (orange band) \cite{Xu:2016hxf} and the lattice QCD results (gray band).
}
\end{figure}

Preliminary results for net-proton fluctuations up to second order at the LHC have been presented recently \cite{Rustamov:2017lio}. In the meantime, the authors of Ref. \cite{Braun-Munzinger:2014lba} have performed an analysis based on the assumption that the lower moments follow the Skellam distribution. In that case it is possible to express the second-order moments in terms of particle yields, for which experimental data exist for a long time. In particular, the analysis was focused on the ratios $\chi_2^B/\chi_2^S$, $\chi_2^B/\chi_{11}^{QS}$ and $\chi_{11}^{BS}/\chi_2^S$. The results for one of these observables are shown in the left panel of Fig. \ref{fig13}. The obtained lower bound on the freeze-out temperature was $\sim$3\% larger than the one obtained from the fluctuation analysis at RHIC.

So far we focused our attention on the fluctuations of electric charge and baryon number, which are largely driven by light flavors (baryon number corresponds to protons and electric charge is pion dominated). The third conserved charge in strong interactions is strangeness. It would be interesting to test whether strangeness freezes out at the same temperature as the other two conserved charges. In particular, ALICE results on particle yields seem to indicate that protons require a smaller freeze-out temperature (compatible with the one obtained from the analysis of fluctuations) than (multi-)strange particles by $\sim18$ MeV \cite{Floris:2014pta}. Among the possible explanations for this, the authors of Ref. \cite{Bazavov:2014xya} have proposed that including in the thermal fit the additional strange states predicted by the Quark Model might reduce this temperature gap. Another idea would be that, due to the large annihilation cross sections, (anti-)proton freeze-out is expected to occur at lower temperatures \cite{Becattini:2016xct,Rapp:2000gy,Rapp:2001bb,Rapp:2002fc,Steinheimer:2012rd,Becattini:2012xb}. Finally, there is the possibility of a flavor hierarchy in the deconfinement phase transition of QCD in which strange quarks, due to their heavier mass, would hadronize at a slightly higher temperature compared to light quarks \cite{Bellwied:2013cta}. Fluctuations of certain particle species are more sensitive than yields to the freeze-out conditions, and might help to resolve this issue \cite{Alba:2015iva}.

Measuring the full strangeness fluctuations experimentally represents unfortunately a remarkable challenge, since it would require the measurements of event-by-event distributions of rare states such as the multi-strange hadrons. Nevertheless, Kaon fluctuations have been presented by the STAR collaboration \cite{Xu:2016hxf}, and even $\Lambda$ fluctuations might become available in the near future. Since the lattice QCD results are usually obtained for conserved charge fluctuations, while the identification of single particle contributions can be challenging, the authors of Ref. \cite{Noronha-Hostler:2016rpd} have proposed a way to isolate kaon fluctuations from lattice QCD simulations. By comparing, within an HRG model framework, the full kaon $\chi_2/\chi_1$ (which takes into account not only primordial kaons but also the products of resonance decays) and the one obtained from primordial kaons in the Boltzmann approximation only, they found that the latter represents a remarkably good approximation of the former. The formula for the charged-kaon $\chi_2/\chi_1$ thus reduces to
\begin{equation}
\frac{\chi_2^K}{\chi_1^K}=\frac{\cosh(\hat{\mu}_S+\hat{\mu}_Q)}{\sinh(\hat{\mu}_S+\hat{\mu}_Q)}.
\end{equation}
As already mentioned, due to the conditions $\langle n_S\rangle=0$ and $\langle n_Q\rangle=0.4\langle n_B\rangle$, the chemical potentials $\mu_S$ and $\mu_Q$ contained in the above formula become functions of $T$ and $\mu_B$, so that it is possible to compare the lattice QCD curve for the above quantity to the experimental measurement to extract the freeze-out parameters for kaons. An example for the highest collision energy at RHIC is shown in the right panel of Fig. \ref{fig13}. Unfortunately, the present uncertainty on the STAR data only allows to extract a lower bound for the kaon freeze-out temperature, $T_f^K\geq$ 145 MeV. 
%{\bf The results on fluctuations are presented. The comparison to experiment is carefully addressed, including all the caveats and things which need to be taken into account. The freeze-out parameters are discussed for electric charge, baryon number and strangeness.}
\section{Transport properties and electromagnetic probes of QCD matter \label{4}}
The transport properties of QCD matter are expected to be severely modified in the vicinity of the phase transition, where the system is strongly coupled. Unfortunately, the observables related to these transport properties are dynamical quantities, very challenging to extract from lattice QCD simulations. On the lattice, it is possible to investigate current-current correlators on a discrete set of points. These correlators $G$ have a spectral representation which involves integrals of spectral functions $\rho$ weighted by the appropriate integration kernels $K$:
\begin{equation}
\hspace{-1.8cm}
G(\tau,\vec{p})=\int_0^\infty\frac{d\omega}{2\pi}\rho(\omega,\vec{p},T)K(\omega,\tau,T)~~~~~~\mathrm{with}~~~~~~
K(\omega,\tau,T)=\frac{\cosh(\omega(\tau-\frac{1}{2T}))}{\sinh(\frac{\omega}{2T})}.
\end{equation}

Extracting the low-frequency and low-momentum limit of such spectral functions, which are the observables of interest, requires the application of inversion methods or a modeling of the spectral functions at
low frequencies in order to integrate over a discrete set of lattice points. In spite of these difficulties, several
results have been obtained recently on the transport properties of matter.

Kubo formulas relate the transport coefficients to the spectral functions: the electric conductivity is related to the light vector spectral function, the heavy quark momentum diffusion coefficient to the color electric correlators and the bulk and shear viscosities to the energy momentum correlation functions.

\begin{figure}[h]
\begin{center}
\includegraphics[width=3in,height=2.6 in]{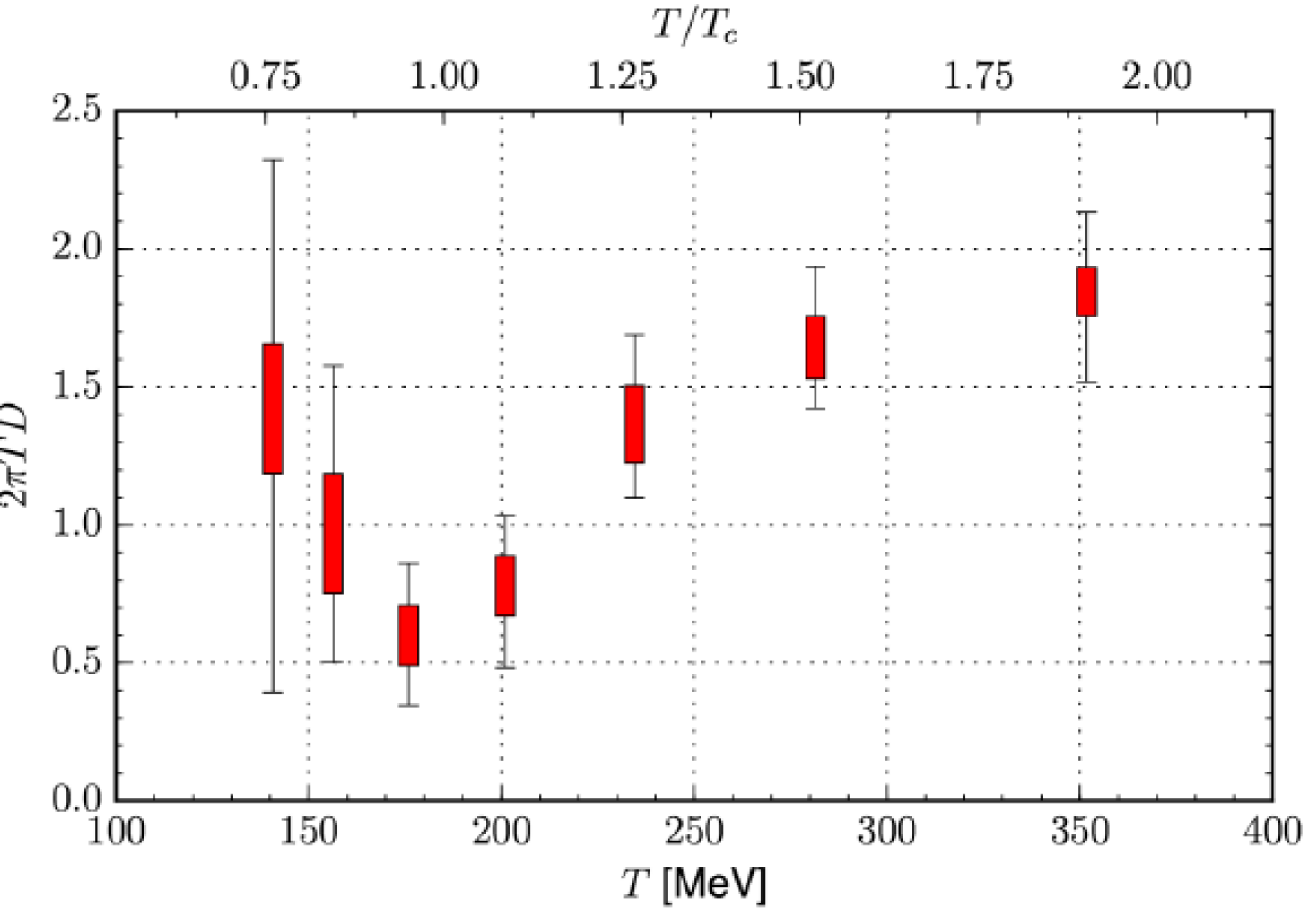}
\includegraphics[width=3.1in, height=2.6in]{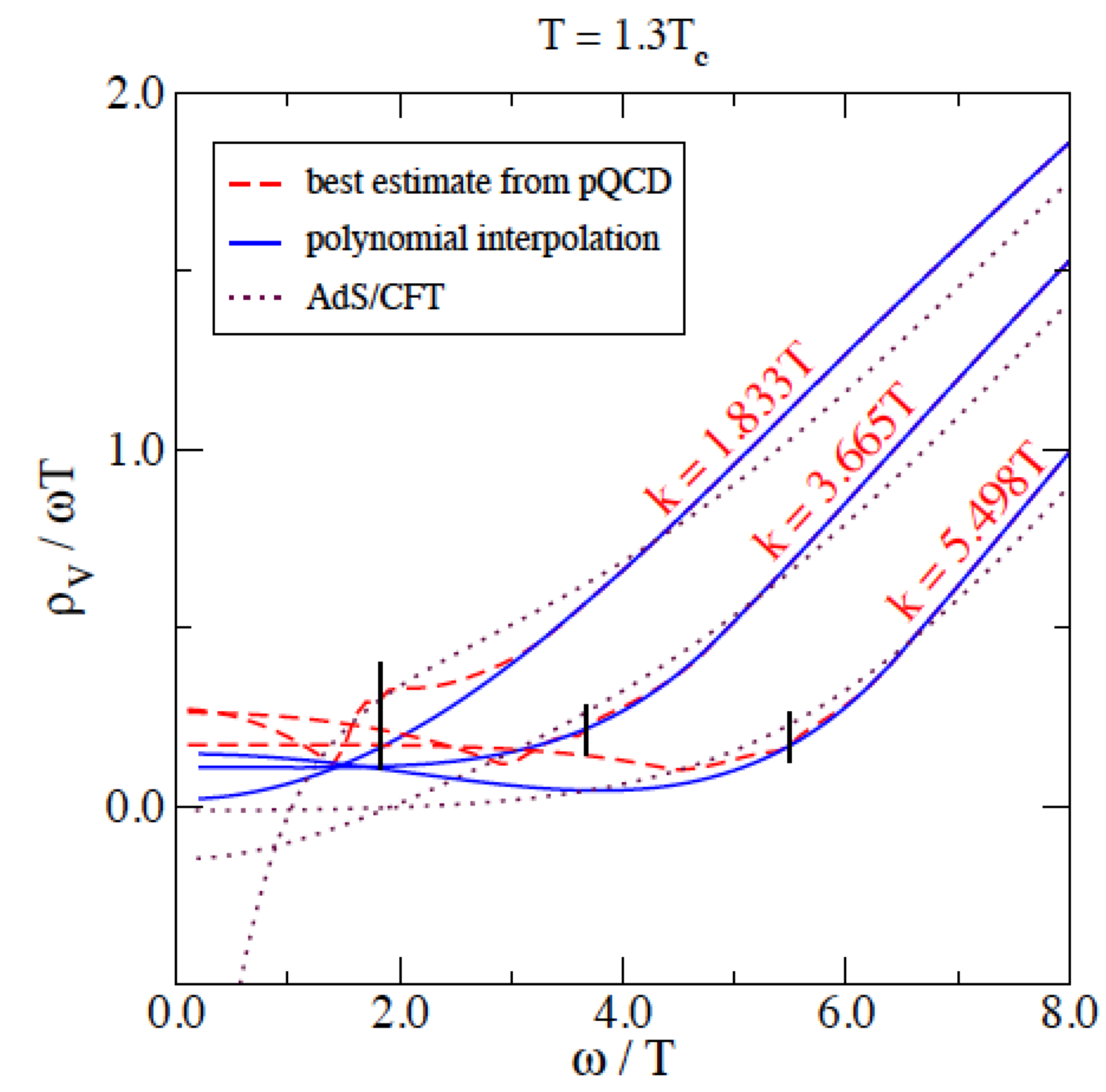}
\end{center}
%\vspace{-.5cm}
\caption{\label{fig14}
Left: From Ref. \cite{Aarts:2014nba}: charge diffusion coefficient $D$ multiplied by $2\pi T$ as a function of the temperature. The vertical size of the rectangles represents the systematic uncertainty due to the uncertainty in the estimate of the conductivity, while the whiskers indicate the statistical jackknife error from both $\sigma$ and $\chi_Q$. Right: From Ref. \cite{Ghiglieri:2016tvj}: spectral functions obtained from the fit of vector meson correlation functions. The AdS/CFT result is from Ref. \cite{CaronHuot:2006te}.
}
\end{figure}

The electric charge conductivity $\sigma$ measures the response of the medium to small perturbations induced by an electromagnetic field. Several recent studies exist for this quantity from lattice QCD \cite{Ding:2010ga,Burnier:2012ts,Brandt:2012jc,Aarts:2014nba}. One of the first studies was performed in the quenched approximation \cite{Ding:2010ga} and the electrical conductivity was obtained at $T\simeq1.45T_c$. An estimate in the case $N_f=2$ was give in Ref. \cite{Brandt:2012jc}. More recently, the authors of Ref. \cite{Aarts:2014nba} have evaluated this quantity in the case of a 2+1 flavor system, albeit with heavier than physical quark masses, by means of the Maximum Entropy Method. The value of $\sigma$ increases by a factor 6 in the range of temperatures between 140 and 350 MeV. The charge diffusion coefficient $D$ has also been obtained, by dividing the electric conductivity by the second order fluctuation $\chi_Q$. $D$ presents a dip in the vicinity of $T_c$, consistent with the expectations of a strongly coupled
system; this result is shown in the left panel of Fig. \ref{fig14}. In this plot, the vertical size of the rectangles represents the systematic uncertainty due to the uncertainty in the estimate of the conductivity from the Maximum Entropy Method, while the whiskers indicate the statistical jackknife error from both $\sigma$ and $\chi_Q$. Other sources of systematic errors, such as the lack of continuum extrapolation and the heavier-than-physical quark masses, are not taken into account.

Thermal photons emitted from the QGP can provide information on the interactions that the partons experience in the plasma, while dilepton spectra in heavy-ion collisions can be related to chiral symmetry restoration. The experimental thermal photon and dilepton rates can also be written in terms of the spectral function in the vector channel \cite{McLerran:1984ay,Moore:2006qn}. Ideally, one would want to compare experimental results to first-principle calculations in a model-independent way. The lattice approach is very challenging, for the reasons explained above. The authors of Ref. \cite{Ghiglieri:2016tvj} have calculated continuum extrapolated vector meson correlation functions at non-vanishing momenta, by combining lattice and perturbative techniques \cite{Ghiglieri:2014kma,Ghisoiu:2014mha,Laine:2013vma} in the regimes where they are under control. In regimes for which lattice results show clear deviations from the weak-coupling prediction, a polynomial description of the spectral shape was used. The resulting spectral functions are shown in the right panel of Fig. \ref{fig14}. From these quantities, the authors estimated the photon production rate at temperatures just above the phase transition. Such results agree with the NLO perturbative predictions of Ref. \cite{Ghiglieri:2014kma} for momenta $k\geq3T$, while for lower momenta the pQCD results represent an overestimate, in apparent qualitative agreement with phenomenology \cite{Burnier:2015rka}.

As for the dilepton spectra, the low-mass dilepton emissivity, which governs thermal radiation from a hot fireball, is directly
proportional to the vector spectral function. Chiral symmetry breaking manifests itself in a splitting of the masses of chiral partners such as the $\rho$ and $a_1$ mesons, which dominates the dilepton signal. Several approaches to this problem exist, based e.g. on QCD and Weinberg sum rules with inputs from lattice QCD for the chiral order parameter, or on hadronic chiral lagrangians used to evaluate the medium effects (for recent reviews see e.g. \cite{vanHees:2007th,Rapp:2009yu,Rapp:2013nxa}). For the $\rho$ meson, the QCD sum rule approach can only provide constraints on model calculations \cite{Hatsuda:1992bv,Leupold:1997dg,Zschocke:2002mn}, whereas the Weinberg sum rule relates the chiral condensate to the energy moments of the difference between the vector and axialvector spectral functions \cite{Weinberg:1967kj}. By combining these two approaches, the authors of Ref. \cite{Hohler:2013eba} found that the $\rho$ meson spectral function which successfully describes the experimental dilepton spectra is compatible with the temperature-dependent quark condensate computed in lattice QCD \cite{Borsanyi:2010bp}. The chiral lagrangian approach is particularly interesting because chiral order parameters can be calculated directly as functions of the temperature. The authors of Ref. \cite{Hohler:2015iba} studied the temperature progression of the $\rho$ and $a_1$ meson spectral functions towards chiral mixing in a pion gas and found a drop of 15-20\% in the chiral condensate at $T\simeq160$ MeV, corroborating the idea that the degeneracy in the spectral functions is coupled to chiral symmetry restoration.

The heavy flavor diffusion coefficient can help to answer the question whether the heavy quarks are relaxing towards local thermal equilibrium in the QGP: it characterizes the movement of the heavy flavor with a momentum of at most the order of the temperature with respect to the medium rest frame. Such a quantity has been estimated on the lattice in the deconfined phase of QCD in Refs. \cite{Ding:2012sp,Francis:2015daa}. Both of these calculations have been performed in the quenched approximation. In Ref. \cite{Ding:2012sp} the heavy quark mass is finite but the continuum extrapolation is missing. The force that the heavy quark feels as it propagates through the QGP is related to a color-electric correlator, which is the main quantity simulated in Ref. \cite{Francis:2015daa} at $T=$1.5 $T_c$. Here, the result has been continuum extrapolated for the first time, but with an infinitely heavy quark mass. The resulting estimate of the heavy quark momentum diffusion coefficient $\kappa$ is a value between $1.8T^3$ and $3.4T^3$, which yields an estimate for the time scale associated with
the kinetic equilibration of heavy quarks $\tau_{kin}=(1.8...3.4)(T_c/T)^2(M/(1.5$GeV)) fm/c. Even though this error bar is relatively large, the analysis contains a realistic estimate of systematic uncertainties. In the non-relativistic limit, $\kappa$ is related to the diffusion coefficient $D$ as $D=2T^2/\kappa$.

\begin{figure}[h]
\begin{center}
\includegraphics[width=3in,height=2in]{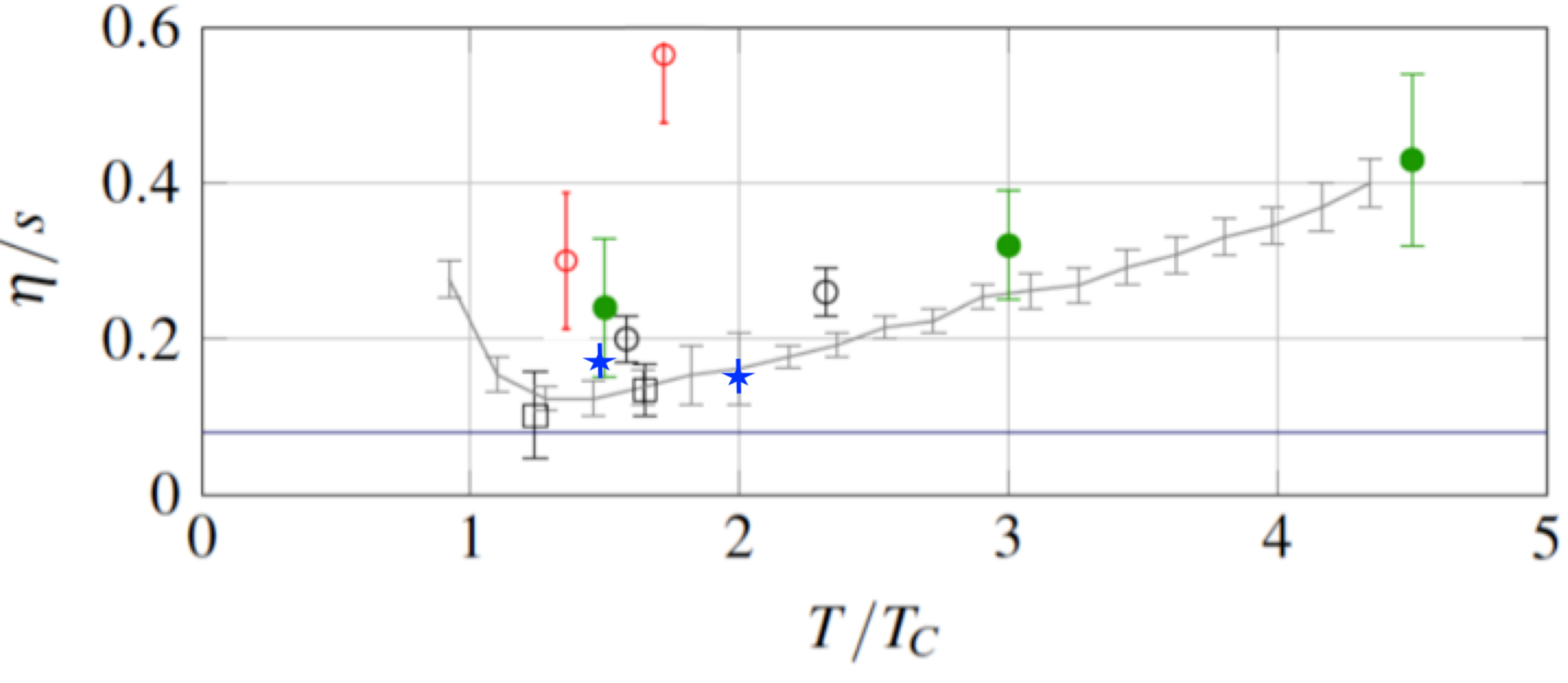}
\includegraphics[width=3in]{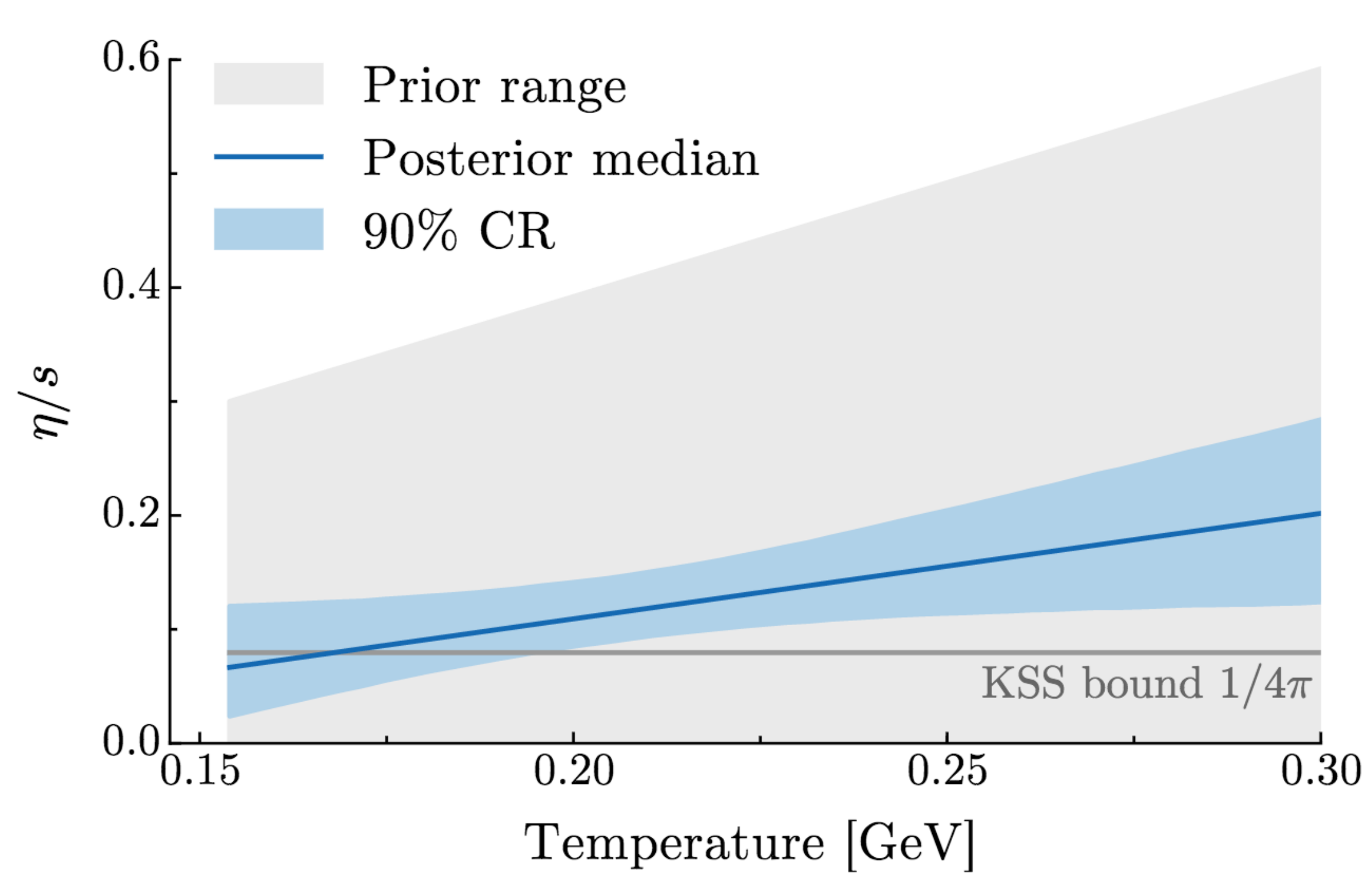}
\end{center}
%\vspace{-.5cm}
\caption{\label{fig15}
Left: Compilation of all available lattice QCD and selected lattice-based results on the pure gauge shear viscosity over entropy as a function of the temperature: H.
Meyer (black squares and circles) \cite{Meyer:2007dy,Meyer:2009jp}, Christiansen et al. (vertical lines)\cite{Christiansen:2014ypa}, Nakamura and Sakai (empty red circles) \cite{Nakamura:2004sy}, S. W. Mages et al. (green full circles) \cite{Mages:2015rea}, A. Pasztor et al. \cite{Pasztor:2018yae} (blue stars). Right: from Ref. \cite{Bernhard:2016tnd}: estimated temperature dependence of $\eta/s$ above the transition temperature from a Bayesian analysis. The gray shaded region is the prior range, the blue shaded band is a 90\% credible region around the median from the posterior distribution (blue line). In both figures, the horizontal line is the bound from AdS/CFT \ \cite{Kovtun:2004de}.
}
\end{figure}

The most remarkable feature of the QGP is its nearly-ideal fluid nature \cite{Teaney:2001av,Teaney:2003kp,Romatschke:2007mq,Song:2008hj}. This unexpected result
relies on the observation that the QGP can be described by nearly ideal hydrodynamics, with a shear
viscosity over entropy density ratio ($\eta/s$) fixed to reproduce the experimentally measured elliptic flow. Due to the successful description of the evolution of the matter created at RHIC and the LHC by means of viscous hydrodynamics, physicists understood that the initial anisotropic pressure gradients in the fireball drive the system to collectively develop a momentum anisotropy. 
Given the observation of elliptic flow at RHIC and more recently at the LHC, the shear viscosity is one of the most important transport coefficients for heavy-ion phenomenology. The gauge-string
duality between Anti-de-Sitter space and conformal field theory has led to the conjecture that for any
relativistic quantum field theory, $\eta/s$ is above the lower bound $1/4\pi$ \cite{Kovtun:2004de}. The ideal fluid behavior of the
QGP has been interpreted as signaling a strongly interacting system, since the value of $\eta/s$ obtained for a
gas of quarks and gluons in a weak coupling regime is an order of magnitude larger. Results on the actual
QCD prediction of viscosity are rather scarce and available only in a purely gluonic system \cite{Nakamura:2004sy,Meyer:2011gj,Mages:2015rea}. The first continuum extrapolated results for this observables in the quenched approximation were presented in \cite{Pasztor:2018yae} at $T=1.5T_c$ and $T=2T_c$; these results are consistent with previous estimates.
For this observable, besides the difficulty
of inverting the energy-momentum tensor correlator, an additional problem arises: the correlator itself is
extremely noisy, and no technique is available to reduce it if quarks are introduced in the simulations. For the quenched case, a powerful solution is the multilevel algorithm \cite{Luscher:2001up}. This algorithm depends crucially on the locality of the action, which
is the reason why so far only quenched results are available. An algorithm which allows to increase the
signal-to-noise ratio is needed to extract this observable also in the full QCD case. Recent attempts to extend the multilevel algorithm to fermions are reported in Ref. \cite{Ce:2016qto}. A compilation of all the results for $\eta/s$ from lattice QCD simulations and a lattice-based approach \cite{Christiansen:2014ypa} is shown in the left panel of Fig. \ref{fig15}. It is important to remark that all these calculations are performed on lattices with rather small
temporal extent, compared to those used in calculations of the electrical conductivity and diffusion constants. 
These sources of systematic uncertainties need to be better estimated.

As already mentioned, recently the Bayesian analysis has been applied to extract the temperature-dependence of $\eta/s$ through a multi-parameter model-to-data comparison. The model is calibrated to multiplicity, transverse momentum and flow data and predicts constraints on the parametrized initial conditions and the temperature-dependence of the transport coefficients in the QGP. The behavior of $\eta/s$ as a function of the temperature is shown in the right panel of Fig. \ref{fig15}, and it is in reasonable agreement with the lattice QCD results for this quantity.

Other theoretical analyses of the experimental results have been performed over the years, aimed at determining this important quantity phenomenologically (for a recent review see e.g. \cite{Shen:2015msa}). In particular, the second-order anysotropic flow coefficient of charged hadrons  measured at RHIC has led to the constraint $\frac{1}{4\pi}\leq(\frac{\eta}{s})_{QGP}\leq\frac{5}{8\pi}$ \cite{Song:2010mg}.
More recently, higher order anisotropic flow coefficients have been measured very precisely \cite{ALICE:2011ab}. This allows one to disentangle the viscous effects from those related to the geometrical shape of the initial state, thus leading to a more precise determination of $\eta/s$: a  phenomenological  extraction  of  $(\eta/s)_{QGP}$ with  a  relative  precision  of  order  5-10\%  now appears  within  reach \cite{Gale:2012rq,Schenke:2014zha}. At the current stage, theoretical models and experimental measurements are both beginning to reach the sensitivity necessary to constrain even the temperature dependence of this ratio  \cite{Shen:2011eg,Niemi:2011ix,Shen:2011kn},  as  well  as  that  of  other  transport  coefficients, such as the bulk viscosity \cite{Noronha-Hostler:2013gga} and various second-order transport coefficients.
%{\bf Discussion and results on some transport coefficients of experimental relevance such as shear viscosity, electric conductivity, heavy flavor diffusion coefficient.}
\section{Heavy flavors and quarkonia \label{5}}
The physics of heavy flavors and quarkonia has reached a new exciting era, during which it is possible to understand the experimental results and relate them directly to lattice QCD simulations \cite{Aarts:2016hap}. Heavy flavors are the ideal probe to study the properties of the QGP: since their mass is much larger than the temperature of the plasma, they are produced in the very early stage of the collision and enable us to test the entire space-time evolution of the system.

The in-medium behavior of quarkonia is usually studied in three different ways:
\begin{itemize}
\item{solve the Schr\"odinger equation for the bound state two-point function, with a $q\bar{q}$ potential simulated on the lattice}
\item{simulate euclidean temporal correlators on the lattice, and reconstruct the quarkonia spectral functions from them}
\item{study the in-medium screening properties of spatial correlators.}
\end{itemize}
The latter represent a complementary way to study in-medium properties of various excitations which is not limited by the finite temporal extent of the lattice. These correlators provide information on the dissolution of bound states \cite{Bazavov:2014cta}, even though they are less easily related to phenomenology.

The first approach is based on the idea that the interaction between the static quarks which form a bound state can be described by an instantaneous, temperature-dependent potential \cite{Mocsy:2008eg,Laine:2006ns,Beraudo:2007ky}. The latter can be obtained from effective theories or lattice QCD results. If the quark mass $m$ is large but finite, quarkonia are expected to survive at temperatures much lower than $m$. One can construct effective theories in which the quark mass is integrated out, and an expansion in $1/m$ is performed. The static limit provides the first term of the expansion, while higher order corrections can be systematically included in the framework of non-relativistic QCD (NRQCD) or potential non-relativistic QCD (pNRQCD), a lower-energy version of NRQCD \cite{Brambilla:2008cx}. It was pointed out in Refs. \cite{Beraudo:2007ky,Brambilla:2008cx} that the effective potential develops an imaginary part which can be related to gluo-dissociation and 
inelastic parton scattering in the medium \cite{Brambilla:2011sg,Brambilla:2013dpa}. The real part of the potential reflects color Debye screening effects. 

\begin{figure}[h]
\begin{center}
\includegraphics[width=6in]{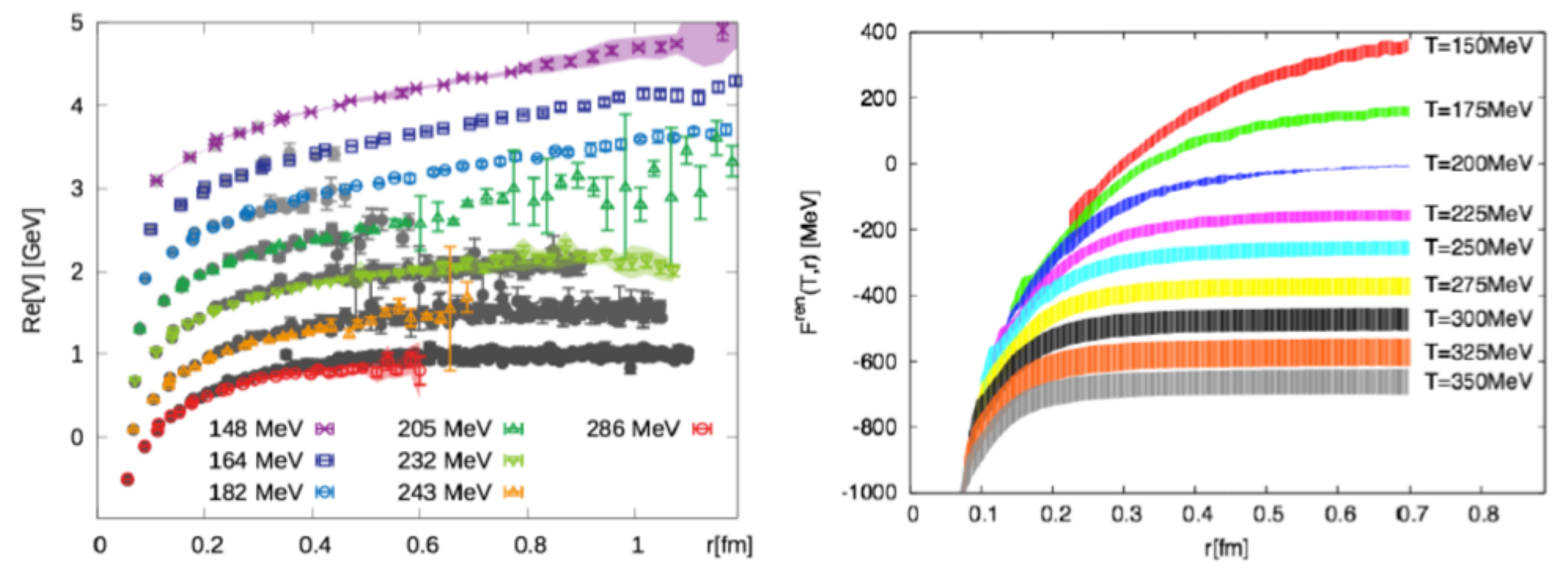}
\end{center}
%\vspace{-.5cm}
\caption{\label{fig16}
Left: From Ref. \cite{Burnier:2014ssa}: the real part of the static interquark potential (colored, open symbols) compared to the color singlet free energies in the Coulomb gauge (full, gray symbols).  Right: From Ref. \cite{Borsanyi:2015yka}: continuum extrapolated results for the 	static $q\bar{q}$ free energy for a system of 2+1 quark flavors at different temperatures.
}
\end{figure}

The complex-valued $q\bar{q}$ potential has been obtained in a system of 2+1 dynamical quark flavors using spectral functions extracted via a Bayesian inference prescription \cite{Burnier:2014ssa}. The real part of the potential lies close to the color-singlet free energy in the Coulomb gauge. These results are shown in the left panel of Fig. \ref{fig16}. Similar results have been obtained by
assuming the validity of the Schr\"odinger equation for charm quarks and extracting the potential from
charmonium correlators directly \cite{Allton:2015ora}. These two results agree with each other and show the typical Debye-screening
flattening of the potential at high temperatures. Recently, the potential method was applied to the study of p-wave quarkonia in Ref. \cite{Burnier:2016kqm}. 

Continuum extrapolated results exist for the $q\bar{q}$ free energy (singlet+octet contributions) obtained from correlators of two Polyakov loops. These results have been obtained for a system of $N_f=$2+1 flavors at physical quark masses \cite{Borsanyi:2015yka}: they are shown in the right panel of Fig. \ref{fig16}. 

The reconstruction of spectral functions from the euclidean correlators is an ill-posed problem: as the correlators are available only for a discrete set of data from lattice simulations, a naive $\chi^2$ fit will lead to a set of spectral functions, all of which can reproduce the discrete set of points for the correlators within error bars. For this reason, additional constraints need to be taken into account, which may narrow down this set of spectra. Such ``priors" are implemented by means of a Bayesian analysis in which a regulator functional is introduced, which helps to select the most probable spectrum.

The Bayesian analysis has been used for reconstructing spectral functions for several years, originally with the Maximum Entropy Method (MEM) \cite{Asakawa:2000tr} and more recently with a Bayesian Reconstruction (BR) approach \cite{Burnier:2013nla}. Currently there are discrepancies in the predictions of the two methods, which can only be resolved towards the Bayesian continuum limit (infinite number of data points and vanishing statistical errors); however, the systematic artifacts of the two methods are now much better understood: while the MEM has been found to be prone to over-smoothing, in particular if only a relatively small number
of data points is available, the BR method can introduce ringing artifacts that may mimic peak
features not actually present in the simulation data.

\begin{figure}[h]
\begin{center}
\includegraphics[width=6in]{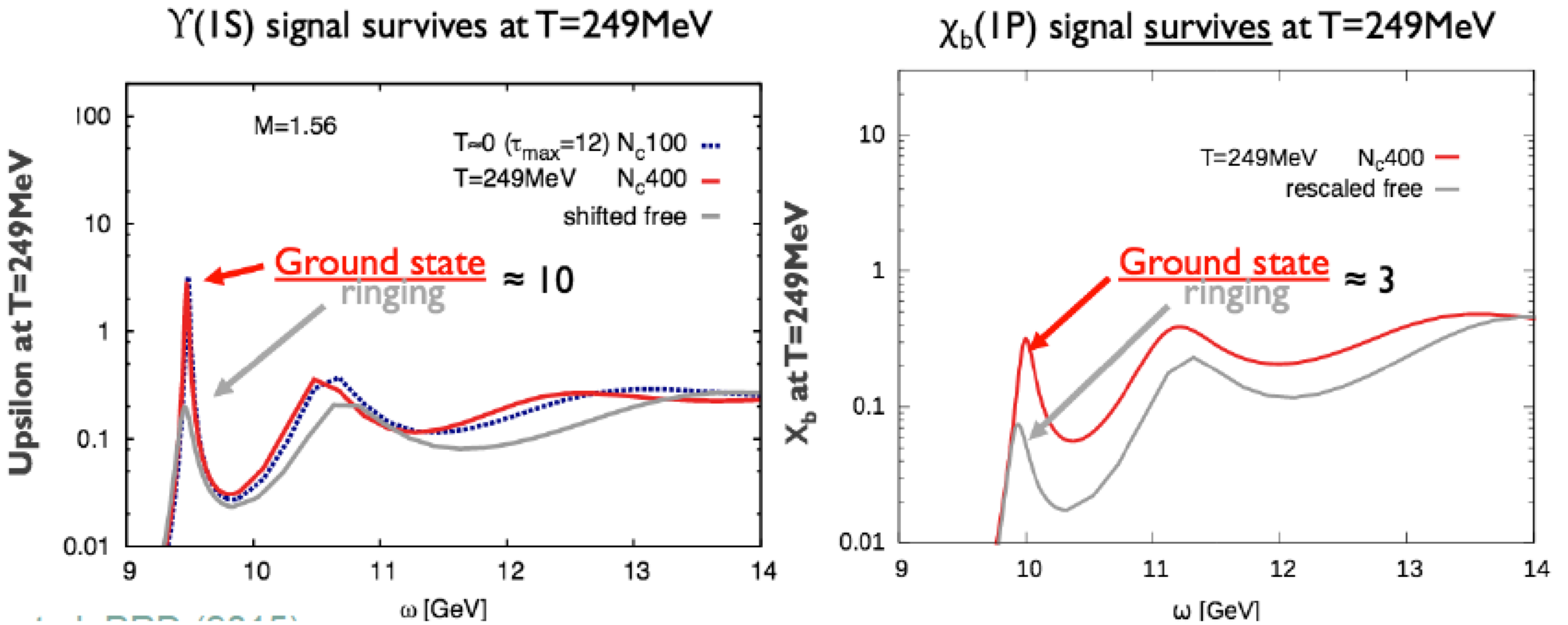}
\end{center}
%\vspace{-.5cm}
\caption{\label{fig17}
From Ref. \cite{Kim:2015rdi}: Spectral functions for the s- (left) and p- (right) wave bottomonium states.
}
\end{figure}

In the case of charmonium spectral functions, both the quenched approximation results and the ones with
dynamical fermions yield a dissociation temperature for charmonium states $T\simeq1.5T_c$ \cite{Ding:2012sp,Aarts:2007pk,Borsanyi:2014vka}. In the case of bottomonium, there is a discrepancy between different analyses. The FASTSUM collaboration, using the Maximum Entropy Method to reconstruct the spectral function, showed that the s-wave state survives up to $T\simeq1.9T_c$, while the p-wave state melts just above $T_c$ \cite{Aarts:2014cda}. By using the Bayesian
method to reconstruct the spectral function, S. Kim et al. find that both s- and p-waves survive in the plasma
up to $T \simeq250$ MeV \cite{Kim:2014iga}: these findings are shown in Fig. \ref{fig17}. To understand this discrepancy, the author of Ref. \cite{Rothkopf:2016rlu} and his collaborators have reconstructed the bottomonium spectral function using both MEM and BR methods and found that this allows to bracket the actual disappearance of the bound state from above and below. In order to work towards the Bayesian continuum limit, the FASTSUM collaboration is generating ensembles with a larger number of points along the Euclidean time direction, while the authors of Ref. \cite{Kim:2014iga} are working on increasing the statistics.

A T-matrix approach has also extensively been used to extract the spectral function of open heavy-flavor and quarkonia \cite{Mannarelli:2005pz,Cabrera:2006wh,Riek:2010fk,Riek:2010py}; this approach consists of using the three-dimensional Bethe-Salpether equation to capture the physics of the in-medium bound states. The concept of a two-body in-medium potential is inherent in this approach. Recently, an in-medium potential has been extracted in this many-body approach \cite{Liu:2015ypa}, defining how finite-width effects (in both potential and heavy-quark propagators) affect the extraction of the underlying interaction kernel. This represents an important, self-consistent step, compared to previous results in which the heavy-quark internal energy and free energy from lattice calculations were used as the input potential in the T-matrix. However, the relationship between this potential and the one discussed above has not been determined yet.

Recently, the authors of Ref. \cite{Mukherjee:2015mxc} proposed an alternative method to study the charm degrees of freedom in the vicinity of the QCD phase transition, based on correlations and fluctuations of conserved charges. Studying the correlations between
charm and baryon number, the contributions to the pressure due to charmed baryons, mesons and quarks can be identified. Besides, by calculating correlators which are equal to each other in the hadronic phase, but differ when charm quarks are liberated, it is possible to identify the onset of deconfinement for charm quarks \cite{Bazavov:2014yba}.

\begin{figure}[h]
\begin{center}
\includegraphics[width=6in]{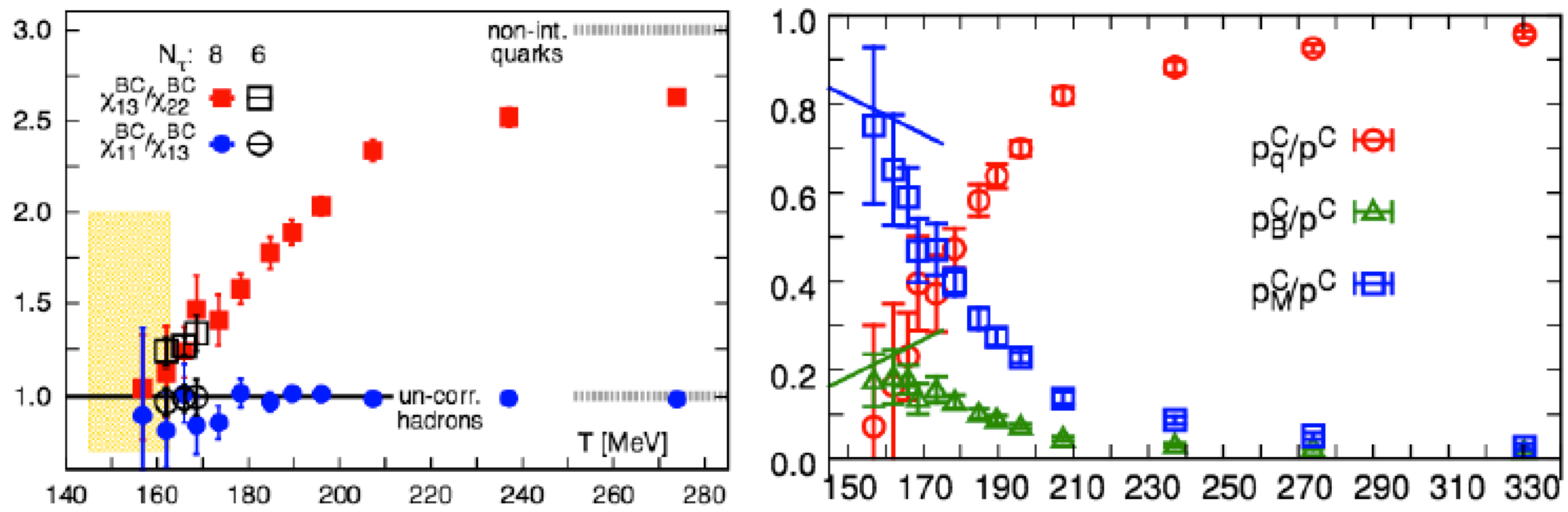}
\end{center}
%\vspace{-.5cm}
\caption{\label{fig18}
Left: From Ref. \cite{Bazavov:2014yba}: onset of deconfinement for charm quarks obtained through the calculation of different Baryon-Charm correlators. Right: From Ref. \cite{Mukherjee:2015mxc}: contribution of charmed baryons, mesons and quarks to the charmed pressure of QCD as functions of the temperature.
}
\end{figure}
This analysis shows that, even if the onset of deconfinement for the charm quark takes place around
$T\simeq$165 MeV \cite{Bazavov:2014yba}, it becomes the dominant degree of freedom in the thermodynamics of the charm sector only
at $T\simeq$200 MeV, while between these two temperatures the dominant contribution to the charmed pressure
is given by open charm meson- and baryon-like excitations with integral baryonic charge \cite{Mukherjee:2015mxc}. The left and right panels of Fig. \ref{fig18} show the onset of deconfinement for charm quarks and the contribution to the pressure of charmed baryons, mesons and quarks, respectively.

One of the main challenges of relating the above approaches to phenomenological models and experimental results is the need to link phenomenological ideas and quantities to theoretically well-defined observables. The phenomenological description of diffusion and jets requires the heavy-quark diffusion coefficient (discussed above) and the jet quenching parameters $\hat{q}$ and $\hat{q}_L$. The latter two enter a phenomenological Fokker-Plank equation based on the assumption that the number density of the hard particles is conserved; this assumption is violated by some processes already at the order $\alpha_s$ in QCD \cite{Ghiglieri:2015ala}. On the lattice it is not possible to isolate such processes, while this can be done in principle in perturbative approaches, in which the contributions of certain ``soft" momentum scales are resummed to all orders. This new effective theory can then be simulated on the lattice \cite{Laine:2012ht,Benzke:2012sz,Panero:2013pla,Laine:2013apa,DOnofrio:2014mld}, leading to a determination of $\hat{q}$. 

Other attempts of relating theoretically well defined observables to phenomenologically relevant quantities include the definition of a temperature-dependent Debye-screening mass $m_D(T)$ \cite{Burnier:2015nsa} and effective coupling constant $\alpha_s(r,T)$ \cite{Kaczmarek:2004gv,Kaczmarek:2005ui,Bazavov:2014soa}, which can be extracted from the heavy-quark potential simulated on the lattice, and used in transport approaches \cite{Song:2016rzw,Song:2015ykw,Song:2015sfa}. Recently, the concept of open quantum systems has been introduced to take into account the possibility that the quarkonium bound state can dissolve already in the regime $rm_D<1$ \cite{Borghini:2011ms,Young:2010jq,Akamatsu:2011se,Akamatsu:2012vt,Rothkopf:2013kya,Akamatsu:2014qsa,Akamatsu:2015kaa,Blaizot:2015hya,Brambilla:2016wgg}; this approach is based on the possibility of clearly separating the constituent quarks from the thermal medium and allows to relate the real and imaginary parts of the in-medium heavy-quark potential to the stochastic evolution of the quarkonium wave function \cite{Akamatsu:2011se}. At the moment, this approach does not yet contain the dissipative effects needed for a consistent thermalization.

%{\bf Available results on the spectral functions of heavy quarkonia. Identification of the charm degrees of freedom in the QGP through fluctuations of conserved charges.}
\section{Conclusions \label{6}}
As illustrated in this manuscript, the quality of lattice QCD results available today allows a direct and meaningful comparison to heavy ion measurements for the first time. Lattice simulations therefore provide a considerable support for the ongoing experimental program at RHIC and at the LHC. The equation of state at $\mu_B=0$ is available since a few years, and the current Taylor expansion coefficients enable us to push these thermodynamic quantities up to $\mu_B/T\simeq2$. The QCD phase diagram is presently known in the same range of $T$ and $\mu_B$, and the current lattice simulations exclude the presence of a critical point for $\mu_B/T\leq2$ in the temperature range 135 MeV $\leq T\leq$155 MeV. The challenge for the future is to extend these results to larger chemical potentials, in order to cover the entire range of the RHIC Beam Energy Scan and the future FAIR and NICA experiments.

Fluctuations of conserved charges can be related to the experimental moments of the corresponding multiplicity distribution. Various comparisons between theory and experiment have been performed in order to extract the freeze-out parameters of heavy ion collisions from first principles. Future perspectives include a better understanding of experimental sources of non-thermal fluctuations, as well as an unambiguous determination of the freeze-out temperature for strangeness. The latter will give us insight into the possibility of a sequential, flavor-dependent freeze-out.

As for the transport properties of QCD matter, extracting dynamical quantities from lattice QCD simulations is challenging because it requires the application of inversion methods or a modeling of the spectral functions at low frequencies in order to integrate over a discrete set of lattice points for Euclidean correlators. The determination of the shear viscosity of the QGP is particularly difficult because the correlator itself is very noisy. Hopefully a new algorithm will be developed, to increase the signal-to-noise ratio in the full QCD case.

Heavy quark physics has reached a new era, in which experimental data can be understood with the help of lattice QCD results. In the case of quarkonium spectral functions, future perspectives include simulating the correlators in full QCD and approaching the Bayesian continuum limit by increasing the statistics and the number of simulation points in the temporal direction.

In conclusion, considerable progress over the last decade has led to reliable finite temperature lattice QCD results. Together with the experimental program and other theoretical approaches, this will help us to achieve a better understanding of bulk and transport properties of QCD matter.
%{\bf Summary and perspective on the synergy between heavy ion collision and lattice QCD.}
%\section*{Acknowledgements}
\ack
The author acknowledges useful discussions with her colleagues and collaborators Paolo Alba, Rene Bellwied, Szabolcs Borsanyi, Zoltan Fodor, Jana G\"unther, Sandor Katz, Volker Koch, Valentina Mantovani-Sarti, Jorge Noronha, Jacquelyn Noronha-Hostler, Paolo Parotto, Attila Pasztor, Israel Portillo, Kalman Szabo. This material is based upon work supported by the National Science Foundation under grants no. PHY-1654219 and OAC-1531814 and by the U.S. Department of Energy, Office of Science, Office of Nuclear Physics, within the framework of the Beam Energy Scan Theory (BEST) Topical Collaboration. An  award  of  computer  time  was  provided by the INCITE program.  This research used resources of the Argonne Leadership Computing Facility, which is a DOE Office of Science User Facility supported under Contract DE-AC02-06CH11357. The author also acknowledges the use  of  the  Maxwell  Cluster  and  the  advanced  support
from the Center of Advanced Computing and Data Systems at the University of Houston.
\section*{References}
\bibliography{thermo}
%\begin{thebibliography}{999}
%\end{thebibliography}
\end{document}